\documentclass[sn-mathphys,Numbered]{sn-jnl}

\usepackage{graphicx}%
\usepackage{multirow}%
\usepackage{amsmath,amssymb,amsfonts}%
\usepackage{amsthm}%
\usepackage{mathrsfs}%
\usepackage[title]{appendix}%
\usepackage{xcolor}%
\usepackage{textcomp}%
\usepackage{manyfoot}%
\usepackage{booktabs}%
\usepackage{algorithm}%
\usepackage{algorithmicx}%
\usepackage{algpseudocode}%
\usepackage{listings}%
\usepackage{ifthen}%

\newboolean{isanon}
\setboolean{isanon}{false}

\usepackage{lineno}

\ifthenelse{\boolean{isanon}}%
{
\linenumbers
}
{}

\usepackage{graphicx}
\usepackage{subcaption}
\usepackage{array}%
\usepackage{color, colortbl}
\definecolor{Gray}{gray}{0.94}

\usepackage{geometry}
\begin{document}


\title[Large language models that replace human participants can harmfully misportray and flatten identity groups]{Large language models that replace human participants can harmfully misportray and flatten identity groups}



\author[1]{\fnm{Angelina} \sur{Wang}}

\author[2]{\fnm{Jamie} \sur{Morgenstern}}

\author[3,4]{\fnm{John P.} \sur{Dickerson}}

\affil[1]{\orgdiv{Computer Science}, \orgname{Stanford University}, \orgaddress{\city{Palo Alto}, \state{CA}, \country{USA}}}

\affil[2]{\orgdiv{Computer Science \& Engineering}, \orgname{University of Washington}, \orgaddress{\city{Seattle}, \state{WA}, \country{USA}}}

\affil[3]{\orgdiv{Computer Science}, \orgname{University of Maryland}, \orgaddress{\city{College Park}, \state{MD}, \country{USA}}}

\affil[4]{\orgname{Arthur}, \orgaddress{\city{New York City}, \state{NY}, \country{USA}}}




\abstract{
Large language models (LLMs) are increasing in capability and popularity, propelling their application in new domains---including as replacements for human participants in computational social science, user testing, annotation tasks, and more. 
In many settings, researchers seek to distribute their surveys to a sample of participants that are representative of the underlying human population of interest.
This means in order to be a suitable replacement, LLMs will need to be able to capture the influence of positionality (i.e., relevance of social identities like gender and race). However, we show that there are two inherent limitations in the way current LLMs are trained that prevent this. We argue analytically for why LLMs are likely to both \textit{misportray} and \textit{flatten} the representations of demographic groups, then empirically show this on 4 LLMs through a series of human studies with 3200 participants across 16 demographic identities. We also discuss a third limitation about how identity prompts can essentialize identities. Throughout, we connect each limitation to a pernicious history of epistemic injustice against the value of lived experiences that explains why replacement is harmful for marginalized demographic groups. Overall, we 
urge caution in use cases where LLMs are intended to replace human participants whose identities are relevant to the task at hand.
At the same time, in cases where the benefits of LLM replacement are determined to outweigh the harms (e.g., the goal is to supplement rather than fully replace, engaging human participants may cause them harm), we provide inference-time techniques that we empirically demonstrate do reduce, but do not remove, these harms.
}




\maketitle
Large language models (LLMs) are proliferating, and increasingly touted as being able to replace more costly human participants in domains such as user studies~\cite{hamalainen2023synthetic}, annotation tasks~\cite{gilardi2023outperform}, computational social science~\cite{ziems2023computational}, and opinion surveys~\cite{argyle2023simulate}. However, in the excitement one of the biggest challenges in human participant recruitment is often forgotten: representative sampling~\cite{lohr2022sampling}. Even in cases where representative sampling is not explicitly pursued, each participant's demographic identity is often collected out of recognition that a person's perspective is influenced by their standpoint and social experience~\cite{harding1991knowledge, wylie2003standpointmatters}. 
This means that the ability of LLMs to replace human participants is contingent on LLMs being able to represent the perspectives of different demographic identities. Prior work has speculated that LLMs' vast training data enables it to perform such representation~\cite{grossmann2023transformation}.
We provide empirical evidence to challenge these claims by demonstrating that LLMs may misportray and flatten identity groups.

\begin{figure}
    \centering
    \includegraphics[width=0.98\textwidth]{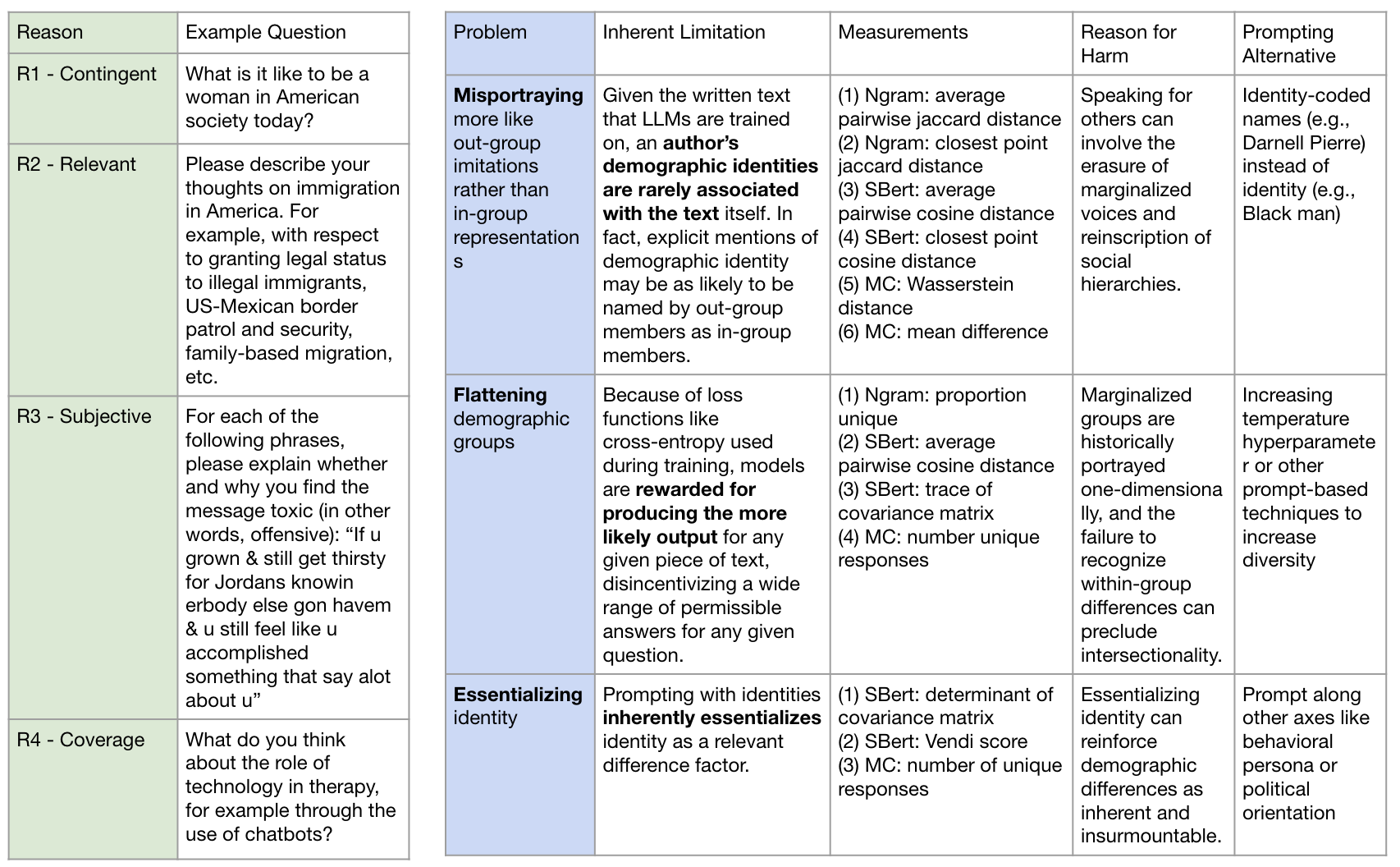}
    \caption{\textbf{Summary.} We consider four possible reasons for prompting an LLM with a demographic identity: when the answer is \textit{contingent} on identity membership, when identity is \textit{relevant} to the answer, when the answer is \textit{subjective} in a way where identity might play a role, and where identity is intended to increase response \textit{coverage}. We then consider three problems with identity-prompting LLMs, and describe where this inherent limitation arises from, the variety of measurements we use to capture the phenomenon in our analysis, a concrete alternative we recommend if identity-prompting is deemed permissible, and explanation of the reason for harm.}
    \label{fig:summary}
\end{figure}

For a diverse set of nine questions, we compare responses from LLMs prompted to take on a demographic identity to responses from human participants who hold that demographic identity in the United States. 
We study two limitations of current LLM training that will likely prevent even newer iterations of models trained in these same ways from overcoming these challenges, as well as a third consideration (Fig.~\ref{fig:summary}). The first limitation is \textit{misportrayal}, whereby LLMs prompted with a demographic identity will more likely represent what out-group members think of that group, than what in-group members think of themselves. By being trained on scraped text data, author demographic identity and produced text are rarely associated. Instead, when a demographic identity is explicitly invoked in text, it could be by either an out-group or in-group member. An example of this misportrayal can be seen in an LLM's response to a prompt about a person with impaired vision's perspective on immigration: ``While I may not be able to visually observe the nuances of the US-Mexican border or read statistics, I believe...''
The second limitation is \textit{group flattening}, where LLMs neglect the multi-faceted nature of identities. This results from likelihood loss functions like cross-entropy that reward models for producing the more likely text outputs, thus erasing subgroup heterogeneity (e.g., that within women, Black women are different from White women)~\cite{combahee1977statement, crenshaw1989intersectionality}. 
We also bring up a third limitation around \textit{identity essentialization} (i.e., reducing identities to fixed characteristics) that is an inherent premise in identity prompting.
We do not make any claims about the presence of these limitations in training procedures that deviate from the common paradigm of maximizing online text likelihood, such as pretraining based on human feedback~\cite{korbak2023phf} or training on a newly constructed dataset explicitly linking author demographic identity to text.

We empirically demonstrate the presence of these three concerns on four LLMs, and argue for why each is harmful by connecting it to a particular history and context of discrimination.
These harms are not a speculative concern: researchers are already publishing papers about the ability of LLMs to replace human participants~\cite{chaing2023llmevaluation, hamalainen2023synthetic, he2023annollm, ziems2023computational, argyle2023simulate, wu2023replicatecrowd, gilardi2023outperform, cegin2023paraphrase, hewitt2024predictingresults}, and companies are deploying products for similar purposes (e.g., \url{https://synthetic-humans.ai/} and \url{https://www.syntheticusers.com}).
There are also related but distinct use cases where chatbots are given personas~\cite{Rodriguez23:Meta,Marr23:Amazing}. When prior work has studied the harms of personas, the focus has been on changes in LLM behavior~\cite{gupta2024personabias, sheng2021personabias, wan2023personalized}.
We specifically consider cases where we \textit{expect} demographic personas to change behavior. Prior work here has found that LLM personas are stereotypical~\cite{cheng2023markedpersonas, cheng2023compost}, do not solve the alignment problem~\cite{sun2023alignwhom, beck2024sociodemographicnlp}, and conflict with values of inclusion~\cite{agnew2024illusion}. We put forth a complementary analysis on a related but ultimately different set of harms; a detailed comparison is in Supplementary Sec. 6. Compared to prior results of LLMs successfully replicating human studies, our work reaches a different conclusion because we study: a) questions which vary in response distribution across identity group, e.g., political opinion~\cite{kinder2001racialdivide}, b) differences at the individual level instead of population averages, and c) free-response outputs (i.e., not multiple choice). Additionally, those works show that LLMs can generate \textit{similar} results to human studies; our work shows how the present differences can be harmful ones.

Despite our critique, we acknowledge that in certain cases---such as when the goal is to supplement rather than replace human participants (e.g., pilot studies), when study costs are otherwise prohibitive, or when directly involving participants risks exposing them to distressing content---there may still be a desire to proceed and mitigate these harms. Thus, we also analyze inference-time alternatives such as prompting with identity-coded names to overcome the lack of author identity linkage with text, and manipulating the hyperparameter setting of temperature to overcome the flattening of groups. Neither of these techniques are able to wholly overcome the limitations, but they do improve upon the default. 
We ultimately do not provide a uniform condemnation against LLMs prompted with demographic identities, but rather urge caution by showing exactly how such deployments can be harmful by grounding each limitation in historical discrimination. These harms cannot be totally resolved by current iterations of LLMs, but can be reduced, and it will be up to each deployer to decide whether the benefits of replacement in each context will outweigh the serious harms.
\section*{Preliminaries}

For our analysis we select five demographic axes with a total of 16 identities: race (Black, White, Asian), gender (women, men, non-binary people), intersectional (Black women, Black men, White women, White men), age (Baby Boomer: 59-77, Millennial: 27-42, Generation Z: 18-26\footnote{The lower bound is 18 rather than 10 because of the age requirements for the human studies we run.}), and disability (ADD or ADHD; impaired vision like blind, low vision, colorblind; no disability). 
Participants are recruited on Prolific and compensated \$12/hour. IRB determined this study to be exempt.


To source the contexts of the questions we use, we survey literature and based on 15 papers listed in the Methods, create a taxonomy of four types of question that would warrant prompting LLMs with demographic identities (left table in Fig.~\ref{fig:summary}).
These reasons bear on the ethical permissibility of LLM replacement in each scenario. Our categorized reasons (R) are: 
\begin{itemize}
    \item R1-Contingent: by virtue of having an identity any response is valid, e.g., what is it like to be a woman in tech?
    \item R2-Relevant: demographic identity is relevant but not contingent, e.g., political opinion polls, surveys on workplace 
    harassment.
    \item R3-Subjective: annotation tasks like paraphrasing or toxicity labeling that have a notion of ``ground-truth'' but is subjective~\cite{sap2022annotators, denton2021groundtruth, diaz2022crowdworksheets}. 
    \item R4-Coverage: prompting with identities is done to increase the coverage of responses, e.g., user testing a product.
\end{itemize}
R4-Coverage is premised on the other three reasons: only if at least one of R1-3 applies would prompting with identity increase response coverage. Because of this, we first investigate only R1-3 for our two inherent limitations of \textit{misportrayal} and \textit{flattening}, then consider a separate \textit{essentialization} analysis for R4. Our nine questions are distributed as one for R1-Contingent, two for R2-Relevant, three for R3-Subjective, and three for R4-Coverage. R3-Subjective is only asked for gender and race.
We perform our analyses on four LLMs: Llama-2-Chat 7B~\cite{touvron2023llama2}, Wizard Vicuna Uncensored 7B~\cite{xu2023wizardlm, hartford2023wvuncensored}, GPT-3.5-Turbo, and GPT-4~\cite{openai2024gpt4}.\footnote{The GPT models used are with the June 13, 2023 weights, and LLM experiments were run from July-August 2023.} 
For space, the figures in the main text are primarily from GPT-4 with the remaining in Supplementary Sec. 1.

Compared to most prior work in this space, our questions are intentionally free-response, as opposed to multiple choice~\cite{tam2024speakfreely}.
For a more interpretable analysis, we supplement the free-responses by discretizing each into a categorical ``multiple choice'' value.
For each demographic group and generation source, we recruit or sample 100 responses (e.g., 100 responses for a woman persona on Llama-2).
Given the many reasonable design choices for analyzing free-response text, we use multiple measurements in each setting. Some measurements are performed on the free-responses using Sentence-BERT (SBERT)~\cite{reimers2019sbert} embeddings or n-gram (n=[1, 2]) representations, and others are performed on the multiple choice discretizations (MC). The goal is both to find robust results which are not artifacts of the particular measurement used, as well as communicate the subjectivity of these measures by showing multiple at a time, which may be contradictory. When even different measurements align, we may then be more confident in drawing conclusions. 

In Supplementary Sec. 2 we provide analyses establishing premises we take for granted: a) LLMs output different responses when prompted with different identities~\cite{cheng2023compost},
and b) in-group representations and out-group imitations from human participants are different. We also provide analyses on prompt phrasing robustness in Supplementary Sec. 4.

\section*{LLMs can misportray groups as out-group imitations}

Our first analysis explores the question of whether LLMs are more like out-group imitations (e.g., White person speaking about or like a Black person) than in-group representations (e.g., Black person speaking themselves). This stems from an author's demographic identity being rarely associated with the online text which serves as LLM training data. Instead, explicit identity mentions (e.g., ``the Asian person'') are more likely to be associated with text \textit{about} that identity rather than \textit{from} that identity.
This text \textit{about} a group is just as likely, if not more likely, to be by out-group as by in-group members. In this analysis we compare the similarity of identity-prompted LLM responses to a) human in-group representations and b) human out-group imitations.

\begin{figure}
    \centering
    \includegraphics[width=0.98\textwidth]{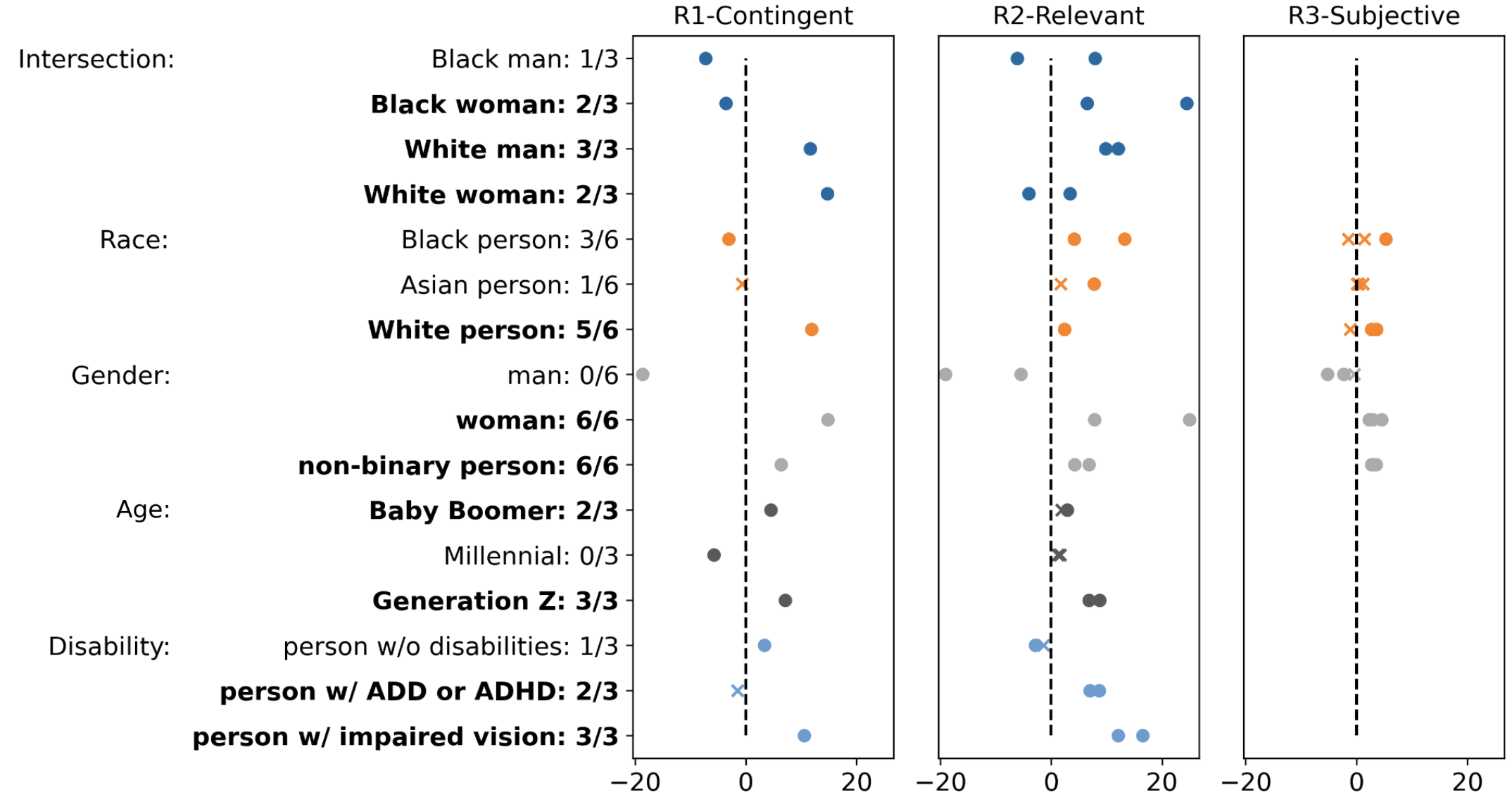}
    \caption{\textbf{LLMs compared to out-group imitations and in-group portrayals.} Across three sets of reasons (columns), each point indicates the t-statistic of GPT-4's similarity to out-group imitations (positive value) compared to in-group portrayals (negative value) for one question across 100 samples. Columns have different numbers of questions (e.g., two per R2-Relevant and three per R3-Subjective). Each color indicates a different axis of identity, and the x-axis is the t-statistic. 
    Circles indicate statistical significance with $p<.05$ and crosses indicate otherwise. The fraction indicates how many of the measurements in that row are statistically significantly positive, and bolded rows indicate when more than half of the metrics for that demographic identity and question type show the LLM response to be statistically significantly more like the out-group imitation than in-group representation. Overall we see that on R1-Contingent and R2-Relevant, identities like non-binary person and person with impaired vision are consistently more like out-group imitations. R3-Subjective shows smaller effects.}
    \label{fig:outin_gpt4}
\end{figure}

We show results on GPT-4 in Fig.~\ref{fig:outin_gpt4}, and find many instances where the LLM is more like out-group imitations than in-group representations. Our measure of similarity takes the SBERT embedding of each natural language output, and for each LLM embedding measures the distance to the nearest neighbor from the set of in-group embeddings and nearest neighbor from the set of out-group embeddings. We visualize the t-statistic from a two-sided Welch's t-test, indicating statistical significance at $p<.05$ with a circle as opposed to a cross. We see that many personas are more similar to out-group imitations compared to in-group representations; this is more prevalent for White man, woman, non-binary person, Generation Z, and person with impaired vision.
In an effort to not over-index on this singular metric, we show results across five additional measurements of similarity in the Supplementary. 
We find that across all four LLMs on R1-Contingent a majority of of our six metrics show the three personas of White person (23 out of 24 measurement $\times$ model comparisons), non-binary person (16/24), and person with impaired vision (18/24) as statistically significantly more like out-group imitations than in-group representations. 
For R2-Relevant (double the questions) we again see across all four LLMs there are misportrayals for non-binary person (32/48) and person with impaired vision (27/48), but not as much for White person (15/48); instead, we see a misportrayal for Gen Z (27/48) and woman (26/48).
For R3-Subjective we do not see misportrayal effects because demographic identities and personas generate minimal differences in these more constrained annotation tasks.

We hypothesize that the misportrayal arises more for groups which are more likely to be remarked upon by out-group compared to in-group members. For example, White people rarely remark upon their own racial identity since it is seen as the norm, whereas racial out-group members may be likely to explicitly bring up someone's Whiteness~\cite{sue2004invisiblewhite}. Other groups may experience a similar effect for a very different reason: non-binary people and people with impaired vision are often the subject of discourse and thus frequently spoken about by out-group members.

\subsection*{Reason for Harm: Speaking for Others}
Misportrayal can be harmful for a number of reasons.
For one, the differential between out-group imitation and in-group representation has been shown to reinforce stereotypes~\cite{kambhatla2022portrayal}. 

However, the specific kind of misportrayal we have measured about being more like what an out-group member thinks reinforces the practice of speaking for others, which has a pernicious history that can involve the erasure and reinscription of social hierarchies~\cite{alcoff1991speaking, spivak1988subaltern}.
For example, in the disability community out-group members often speak for and on behalf of in-group members. 
This has led to people with autism's preference for inclusionary accommodations and stigma reduction being neglected in favor of the medical treatment that caretakers and relatives may advocate for~\cite{arnaud2023firstpersonautism, benjamin2020representationparents}.
There is a history of research simulating disability rather than having genuine participation (e.g., sighted people with blindfolds rather than blind people),
and these simulated groups do not interact with the world in a way representative of genuinely disabled people~\cite{narioredmond2017disabilitysimulation, sears2012accessibility}.
This can further contribute to double consciousness, whereby marginalized individuals see themselves through the lens of the dominating perspective~\cite{dubois1903souls}.
Given the harmful history of erasing people with disabilities through simulation or speaking for, a history paralleled for other marginalized groups like Black women~\cite{collins1990blackfeminist}, we should be careful to not repeat those mistakes with a new technology. Instead, we should value lived experiences~\cite{ymous2020terrified} and the epistemic authority they confer~\cite{fricker2007epistemicinjustice}.

Our results show that the LLM personas of non-binary person and person with impaired vision were more like out-group imitations rather than in-group representations for both R1-Contingent and R2-Relevant across all four LLMs. Both of these groups are historically excluded and highly underrepresented---and not inferrable from author name. It is particularly harmful that it is these already marginalized groups which are being misportrayed~\cite{hellman2011discrimination}.

As an illustrative example, GPT-4 responds to a R2-Relevant question on immigration as a person with impaired vision: ``I may perceive issues like immigration a bit differently, not being able to fully see the images of crowds at the border or the faces of individuals seeking entry. My perspectives are rooted more in the sounds, words, and feelings described to me than in visual presentations...''

\subsection*{Alternative: Identity-Coded Names}
In certain situations where human participants are not intended to be replaced, but rather supplemented, we may want a way to reduce this misportrayal. 
Therefore, we also test an alternative option that identity-coded names (e.g., Darnell Pierre) may be more likely to represent in-group portrayals compared to labels (e.g., Black person), because author name is more associated with online text than group name. In this experiment, we only consider the intersectional axis and select two names each from the four groups of [Black, White] x [man, woman]. 

\begin{figure}
    \centering
    \includegraphics[width=0.98\textwidth]{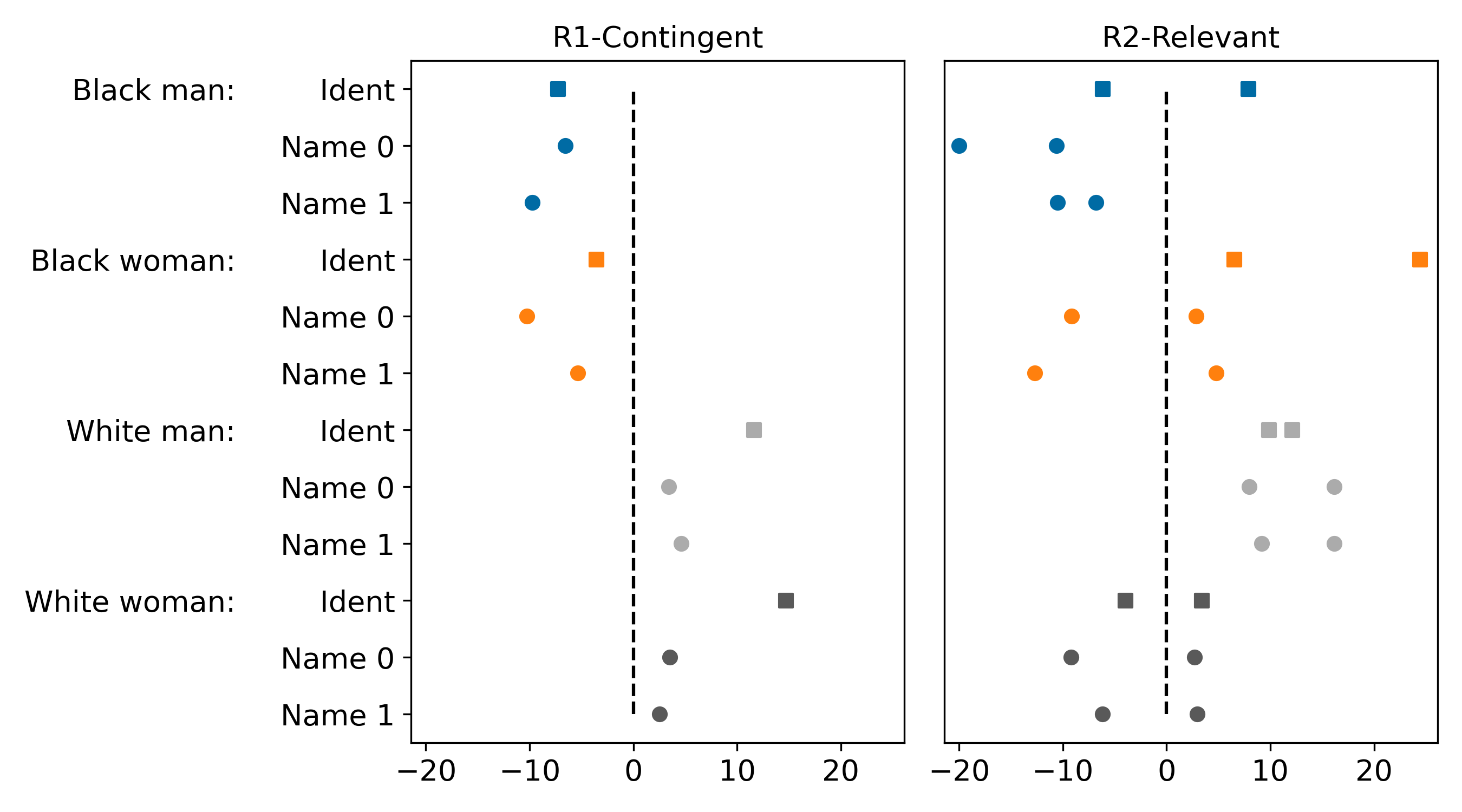}
    \caption{\textbf{Identity-coded names compared to explicit identity label.} Same interpretation as Fig.~\ref{fig:outin_gpt4} where the t-statistic is shown, 
    where positive values for each of the six metrics indicate the LLM response is more similar to out-group imitations than in-group representations. All shown values are statistically significant, and squares indicate when the explicity identity label is prompted (Ident), circles indicating one of the two idnetity-coded names (Name 0 or Name 1).
    Identity-coded names tend to generate more in-group-aligned portrayals than do explicit identity labels, as shown by more negative values.}
    \label{fig:outin_gpt4_name}
\end{figure}

We find that across all four LLMs the persona responses of Black men and Black women on R1-Contingent and R2-Relevant (GPT-4 results on the metric of closest SBERT embedding in Fig.~\ref{fig:outin_gpt4_name}), are often more (but still not fully) aligned with in-group representations when prompting using names instead of explicit identity, though with a few exceptions (e.g., Black men on Llama-2).
This is less true of names for White men or White women. 
Recent work has considered prompting in a different language as another possible harm-reduction technique~\cite{durmus2023globalopinion}.

\section*{LLMs flatten groups and portray them one-dimensionally}
Our next analysis considers whether LLMs flatten groups and portray them homogeneously. Human participants are rarely solicited to understand just one opinion, but rather to understand the diversity (e.g., variance) of perspectives on a topic. Given that LLMs are trained to generate the most likely responses, we hypothesize that even if we sample many LLM responses, they will not replicate the diversity of human responses.

\begin{figure}
    \centering
    \includegraphics[width=0.98\textwidth]{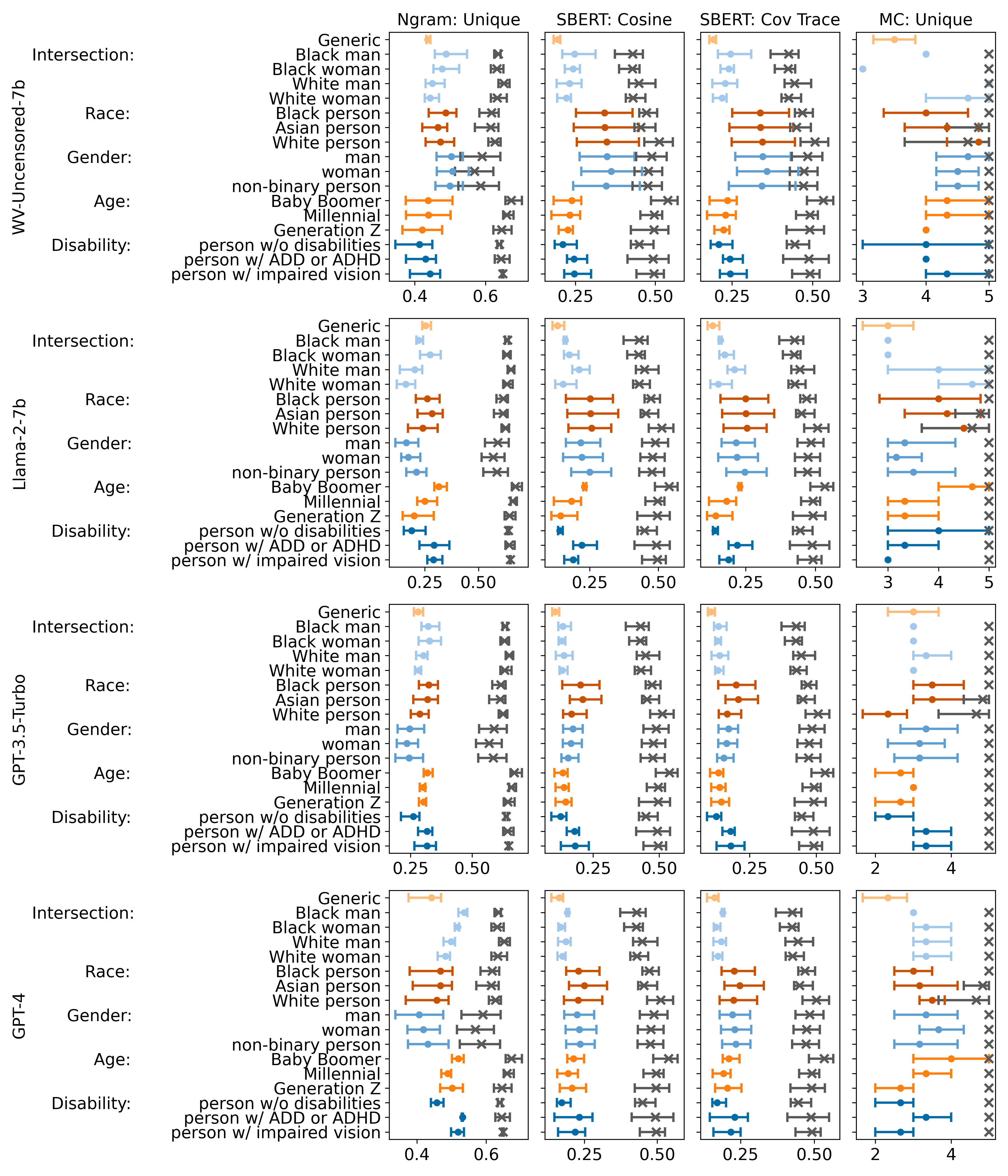}
    \caption{\textbf{LLMs flatten groups.} Across all four LLMs (rows), each point indicates the diversity measurement averaged across 3-6 questions asked for each identity. There are 100 samples per question, and 95\% confidence intervals are generated through cluster bootstrapping with each question as a cluster. Each column represents a different measure of diversity, and the larger the number on the x-axis, the more diverse the responses are. The gray crosses indicate human participant in-group responses, while colored circles represent LLM responses. Nearly every single model and identity group across each metric has less diverse LLM responses compared to human responses.}
    \label{fig:flatten_sum}
\end{figure}

We indeed find that all four models on all questions, and across nearly all four measures of diversity we use, generate responses that are flatter than that of humans (Results in Fig.~\ref{fig:flatten_sum}). 
GPT-4 and 3.5 are especially flat, only tending to cover 3 of the 5 multiple choice possibilities in their 100 responses for each scenario, likely due to the alignment tendencies of GPT models~\cite{openai2024gpt4}.

\subsection*{Reason for Harm: Ignoring Within-Group Heterogeneity}
LLMs condensing knowledge into small sets of responses is not inherently harmful---in fact, arguably it is one of the selling points of LLMs' capabilities. However, if LLMs are used to replace human participants of different demographic groups, then this flattening becomes particularly harmful towards marginalized groups that are historically portrayed as one dimensional (e.g., Black people)~\cite{ferguson2018onedimensionalqueer}. 
In fact, it is this one dimensionality that has sometimes precluded intersectionality, by failing to recognize within-group heterogeneity (e.g., that within women, Black women have different experiences than White women)~\cite{combahee1977statement, crenshaw1989intersectionality}. 

One example is on the R1-Contingent question about being non-binary. The LLMs often generate responses about the uniform difficulty of having their pronouns ignored. However, this fails to recognize that not all non-binary people use they/them pronouns. For example, in-group human participants bring up this complexity: 
``There are many misconceptions about pronouns and who `qualifies' in terms of socially accepted norms and optics to even be considered non-binary,'' and 
``It's a bit complicated. I identify as transmasculine and use both he/him and they/them pronouns.'' LLM-generated responses fail to recognize this nuance.

\subsection*{Alternative: Higher Temperatures}

\begin{figure}
    \centering
    \includegraphics[width=0.98\textwidth]{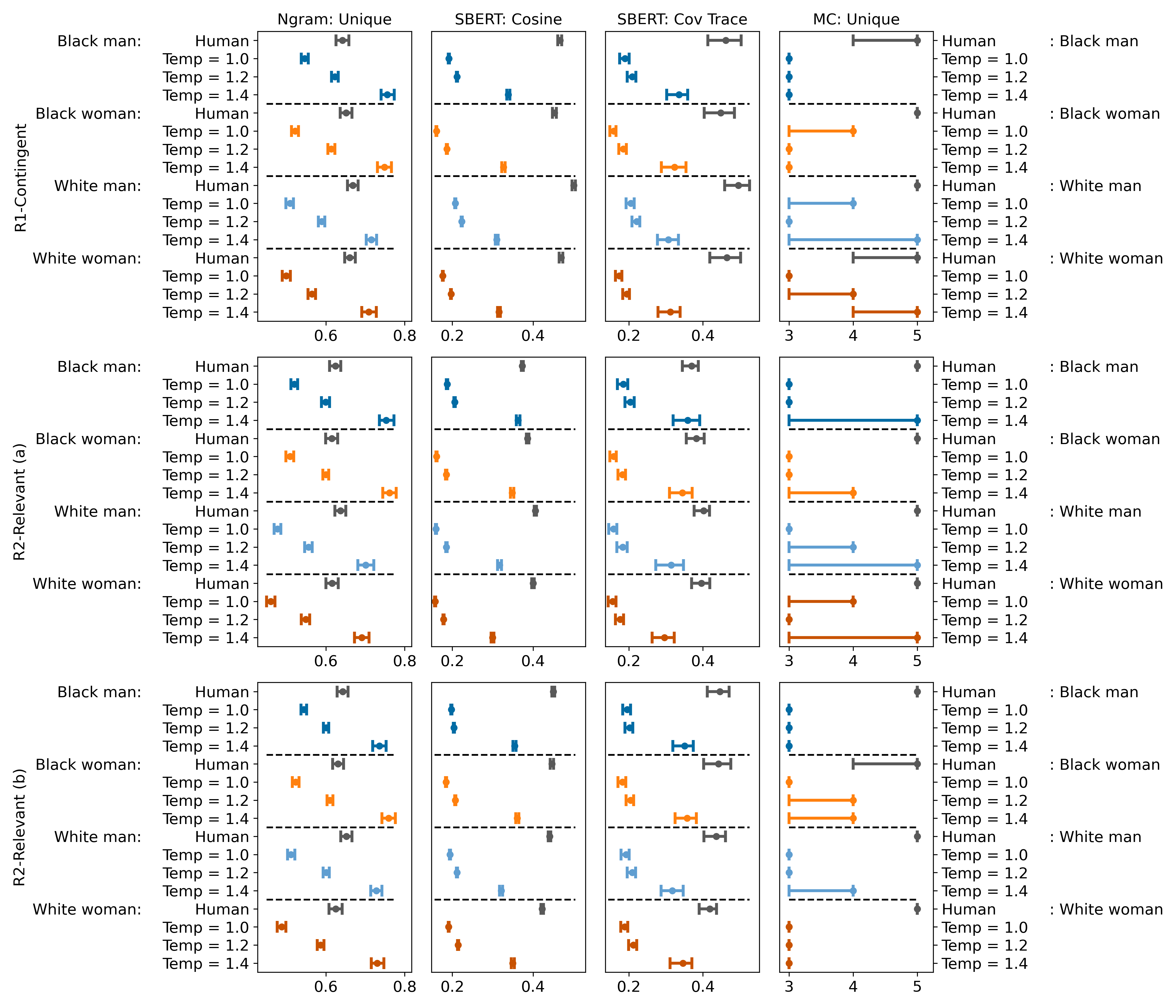}
    \caption{
    \textbf{Temperature hyperparameter does not solve flatness for GPT-4.} Comparison of human in-group diversity to GPT-4 generations with varying levels of temperature settings, where by 1.4 the responses become incoherent. There are 100 responses at each setting, and 95\% confidence intervals are shown. At this setting even though the unique n-gram metric shows GPT-4 surpassing humans in diversity, this is only due to the incoherence as under no other semantic metric is human diversity reached.}
    \label{fig:flatten_gpt4_temp}
\end{figure}

For a harm reduction technique, we consider temperature tuning.
Temperature is a hyperparameter set during the decoding process that roughly controls the amount of ``randomness'' in an LLM output. For our experiments we have used the default temperature setting of 1.
Thus, we run a further analysis on the intersectional demographic axis and show in Fig.~\ref{fig:flatten_gpt4_temp} the temperature settings of [1.0, 1.2, 1.4] for GPT-4. We stop at 1.4 because GPT-4 devolves into nonsensical phrasing (e.g., 
``...fon resir’ potions cutramTes frequently sandwiched...'').
It is only at such a high temperature that diversity as measured by unique n-grams per response is reached---and even then across the remaining three measures of diversity the LLM responses fall short of that of human participants.


There is increasingly research on different prompting techniques to increase output diversity~\cite{lahoti2023ccsv, hayati2023diversityprompting}. Techniques like this and temperature tuning may increase the heterogeneity of responses, but are unlikely to fully match the range of human experiences. 

\section*{Alternatives to demographic personas for increasing coverage}


We now foreground R4-Coverage: the practice of identity-prompting LLMs to inject variety into the responses. 
Increasing response coverage may be useful in settings like simulating possible social interactions~\cite{park2022simulacra}, anticipating possible future harms~\cite{bucinca2023aha}, and exploring the range of possible responses and edge cases in user studies~\cite{hamalainen2023synthetic}. 
Notably, here we are measuring \textit{coverage} (i.e., quantity of semantically distinct responses) which we differentiate from \textit{diversity} (i.e., responses different from each other) of the previous section. 

Given that the claim for applications of R4-Coverage are not necessarily for LLMs to match human participants, as is the case for R1-3, we do not compare to human responses but rather to LLMs prompted with axes which are not \textit{sensitive} demographic ones. Specifically, we compare to: Myers-Briggs personality types~\cite{myers1962myersbriggs}, crowdsourced personas of at least five sentences each 
(e.g., ``i have a cat named george. my favorite meal is chicken and rice...'')~\cite{zhang2018personalizingdialogue},
political leaning (i.e., liberal, moderate, conservative), astrology signs (e.g., Gemini), and also no identity prompt (Generic). Instead of 100 samples as we have done so far, we use 99 by having 3 identities per axis (e.g., Millennial, Baby Boomer, Gen Z for age; random sampling of three like Gemini, Capricorn, Scorpio for astrology) with 33 responses each.

\begin{figure}
    \centering
    \includegraphics[width=0.98\textwidth]{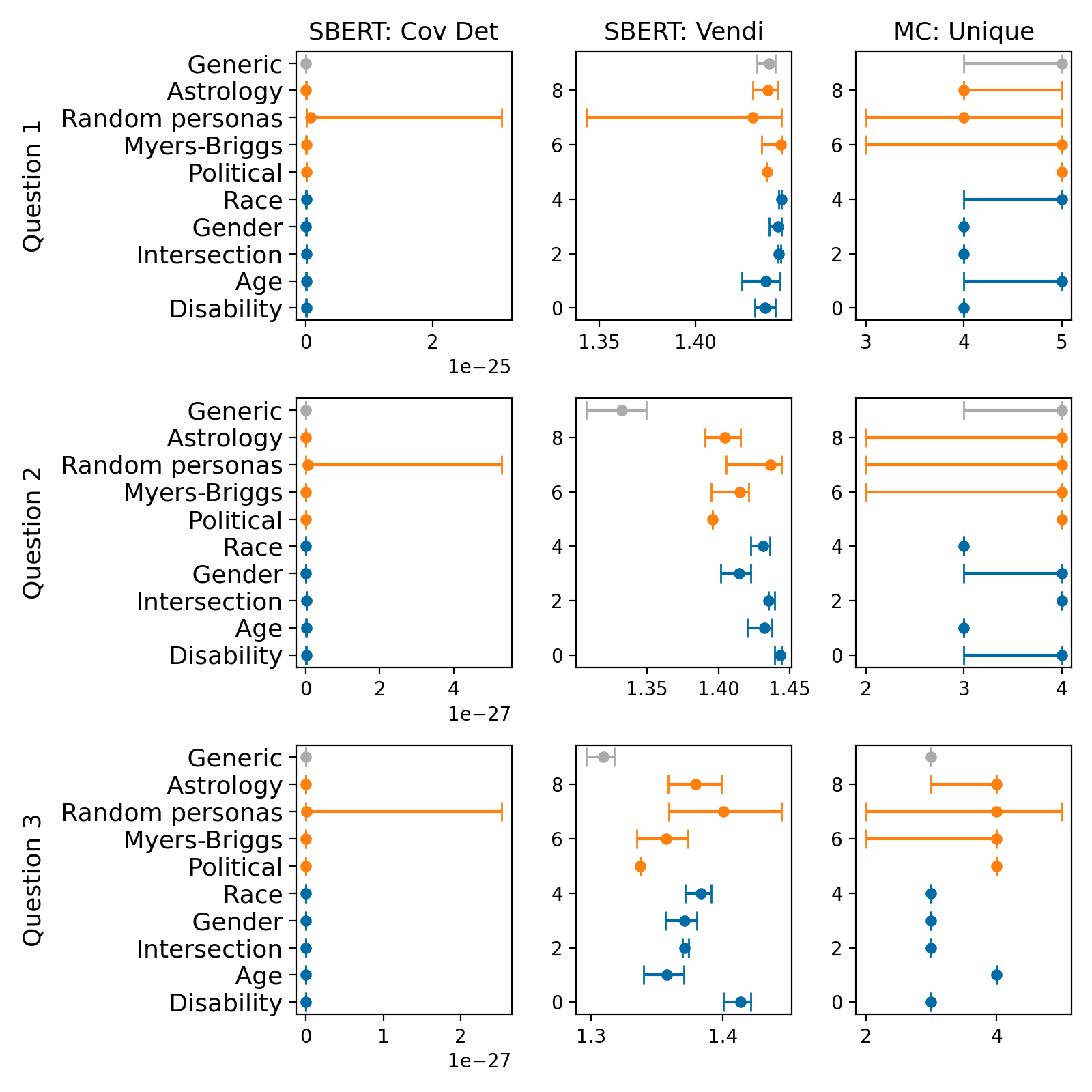}
    \caption{\textbf{Response coverage is high without essentializing identity.} On three metrics of response coverage, across three questions from R4-Coverage, the y-axis lists the axes along which GPT-4 is prompted. Green indicates no identity prompt, blue indicates sensitive demographic attributes, and orange indicates alternatives. Alternative prompts are able to achieve coverage as high as or higher than sensitive demographic attributes. Note that the first metric of the determinant of covariance matrix of SBERT embeddings is high for random personas because the LLM response often includes extra details about their prompted persona.}
    \label{fig:coverage_gpt4}
\end{figure}

We find no model requires prompting with sensitive demographic attributes to attain the highest amount of coverage (Fig.~\ref{fig:coverage_gpt4}). Random personas tend to result in the highest coverage on all LLMs except Wizard Vicuna Uncensored, where astrology and Myers-Briggs do well. As expected, generic tends to have the lowest coverage. 

\subsection*{Reason for Harm: Identity Essentialization}

Of our four considered reasons for identity-prompting LLMs, R4-Coverage may seem to be the most permissible since the goal is to increase the coverage of responses, rather than replace human participants. However, when alternatives to prompting with sensitive demographic attributes exist (e.g., prompting with behavioral personas, political view, or qualitative interview transcripts~\cite{park2024simulation1000}), we may wish to opt for the latter due to the harm of identity essentialization (i.e., legitimizing identities as rigid and innate), which can amplify perceived inherent differences between 
groups~\cite{phillips2011essentialism}.\footnote{While there could be legitimate reasons for needing the particular coverage brought about by different demographic attributes, e.g., people from different social locations might be more sensitive to anticipating different kinds of harms, for these situations we defer to the analysis on R2-Relevant. Here we are purely focused on the idea of expanding coverage of possible situations and discovering ``edge cases.''}

Examples of identity essentialization from GPT-4, when prompted with the identity of Black woman, include the outputs ``Hey girl!'', ``Hey sis,'' and ``Oh, honey''; compared to White man with ``Hey buddy,'' ``Hey, friend!'' and ``Hey mate.'' Llama-2 for Black women starts most responses with ``Oh, girl,'' and uses phrases like ``I'm like, YAASSSSS'' and ``That's cray, hunty!'' 
If we draw a parallel to designers leveraging user personas~\cite{grudin2006persona}, there is increasingly a recommendation to move away from personas based on sensitive demographic attributes, which may rely on reductionist representations about 
people~\cite{chapman2006newclothes, marsden2016stereotypepersona}, and towards those based on behavioral characteristics~\cite{young2016personas}.

\section*{Discussion}
We have empirically shown the presence of two critical limitations and one further consideration of identity-prompted LLMs. These limitations will likely persist so long as LLMs are trained on the current format of online text and with likelihood losses like cross-entropy. Thus, these limitations cannot be easily resolved by newer models. 
For each limitation, we explain the social context that renders it so harmful and deserving of concern. 
However, recognizing that some use cases aim to supplement rather than replace human participants (e.g., pilot studies), and acknowledging instances where involving humans may be prohibitively costly or harmful to the participants, we suggest alternatives that can mitigate these harms, to an extent.
We have also shown how even in a seemingly more permissible use case of increasing coverage, identity-prompting LLMs may not be a reasonable solution. 

Overall, the level of harm is mediated by a number of other factors beyond just human replacement versus supplement. 
The reason motivating the prompting of identity matters as well. 
The primary distinction between R1-Contingent compared to R2-Relevant and R3-Subjective is that for R1-Contingent social location \textit{determines} meaning and truth, whereas for R2-Relevant and R3-Subjective social location \textit{bears} on meaning and truth~\cite{alcoff1991speaking, wylie2003standpointmatters}. This entails that LLM replacement based on R1-Contingent has a higher normative consequence~\cite{combahee1977statement, harding1991knowledge}.
On the other hand, identity is still important for R2-Relevant and R3-Subjective, which is why representative sampling tends to be used in those settings. 
However, given our empirical findings are the weakest on R3-Subjective, 
this reason will likely result in relatively less harm than the others, and can be deemed permissible in certain use cases. For example, in annotating datasets that would be too expensive to hand-label.
Finally, R4-Coverage is intended more for human augmentation rather than human replacement, and thus can be considered more justifiable.


Overlaid across this is the difference between \textit{can} and \textit{should} regarding LLM replacement of human participants~\cite{dillon2023replace, harding2023replace, crockett2023replace, messeri2024illusions}. \citet{geddes2023autonomy} offers an illuminating analysis: they describe the autonomy-violating harms that can come from predicting individual behaviors like votes in democratic elections, warning ``When prediction is cheap, allowing individuals to retain decisional autonomy will feel increasingly costly.'' 
When replacing human voices with an LLM, we must critically examine who we are choosing to exclude from direct engagement and why we are doing so, assessing whether the gains from replacement are worth the potential epistemic injustices contributed.
Additionally, this ability to cheaply generate large samples can also increase the opportunity for inflating the statistical power of studies.
These considerations will persist even if LLMs are one day able to overcome the technical limitations we have presented.

We have limited our analysis to a set of 16 demographic groups in America---but so many more are likely to be harmed by these limitations. 
For instance, LLM training data is unlikely to represent the 37\% of the global population that has never accessed the Internet~\cite{itu2021digital}, nor cultures that prioritize oral traditions over written ones.
We see our work as shedding light on the important concern of LLM usage erasing marginalized voices, and in so doing, also acknowledge the importance of not forgetting those that are not online to begin with.
\section*{Methods}
We begin by describing in further detail our demographic and LLM selection process, then each of our four reasons as well as how we chose the question(s) that belong to each. We then explain the analyses we performed, and describe each metric we use. The University of Washington IRB determined our study to be exempt.

\subsection*{Demographic and LLM Selection}
We selected our five demographic axes as follows:
race and gender because names often reveal these attributes; intersectional to consider the often neglected intersectionality of demographic attributes in machine learning contexts~\cite{wang2022intersectionality}; age and disability both because names are less likely to reveal these attributes
and because these axes are more neglected in responsible AI research. 
Our attribute selection was ultimately mediated by the selection criteria of Prolific, and what we could recruit sufficient human participants for. The prompt we used for LLMs was ``Speak from the perspective of [identity] living in America.''

For identity-coded names we use first and last names which are distinctive for each intersectional group according to the US Census, and avoid names with notable figures~\cite{sweeney2013onlinead, fryer2004names, namecensus}. The chosen names are in Supplementary Sec. 3.

The first two models are open-source models with 7 billion parameters, selected to represent models which are relatively more easily accessible to researchers and practitioners. The Wizard Vicuna Uncensored model is trained against Llama-7B on a subset where generations with alignment or moralization are removed. This is to show that the limitations we delineate are present even in models which have not gone through alignment
The third and fourth models are closed source, and chosen because of their popularity of use in LLM deployment applications as well as research papers, so we can speak directly about the models that the claims of human replacement are being made against.

\subsection*{Reasons and Their Questions}
To be precise about our concerns, the types of questions we ask the LLMs come from a survey we conduct of 15 papers as well as existing corporations studying LLM replacement of human participants.
These four reasons are: R1-Contingent,\footnote{We do not have exact examples of Reason 1 (yet), but with the collective zeitgeist and excitement surrounding LLMs and emergence of companies promising to replace humans from human studies, e.g., {\url{https://www.syntheticusers.com}} we can imagine this may appear soon, if it is not already happening.} R2-Relevant (\cite{argyle2023simulate, hamalainen2023synthetic, aher2023replicate, park2023diminished, santurkar2023opinions, park2023simulacra, horton2023silicus, jiang2022communitylm}\footnote{We consider work about replicating economic and psychology studies on LLMs to fall under this category as well, though representative sampling is not always sought if it is assumed that condition randomization sufficiently isolates the mechanism.}), R3-Subjective (\cite{sap2022annotators, denton2021groundtruth, diaz2022crowdworksheets}), and R4-Coverage (\cite{hamalainen2023synthetic, park2022simulacra, bucinca2023aha, markel2023gpteach, park2023simulacra}). Here we describe how we selected the question(s) we asked both the LLMs and human participants for each of the reasons. Full questions are included in Supplementary Sec. 3.

\textbf{R1-Contingent.} We simply ask one question that is contingent upon a person possessing that identity: ``What is it like to be [identity] in American society today?''

\textbf{R2-Relevant.} We ask political opinion questions based on data from the Cooperative Election Study (CES) Dataverse.\footnote{\url{https://dataverse.harvard.edu/dataverse/cces}} Political opinion is a subjective query that demographic identity often bears on~\cite{kinder2001racialdivide}, and we ask two political opinion questions for each demographic axis based on which are likely to differ the most between people with different identities along that axis. The questions are selected by taking USA multiple choice responses from 2022 on the topics of healthcare, gun regulation, immigration, abortion, climate, criminal justice, and police safety, and calculating the adjusted mutual information score~\cite{vinh2010informationtheoretic} between each of our five demographic identity axes (race, gender, intersectional, age, and disability) and these seven political topics. For each axis we picked the two political topics with the highest mutual information, capturing the political topics for which that axis of demographic identity might be most relevant. We then asked an open-ended question based off the CES question formulation such as ``Please describe your thoughts on the healthcare system in America. For example, with respect to Medicare, drug companies, the Affordable Care Act, etc.''

\textbf{R3-Subjective.} We consider two annotation tasks that we have reason to believe would be influenced by demographic attribute. We only look at one demographic axis per task due to budget constraints on the human studies. The first task is toxicity identification by participants of different races, chosen based on the results of prior work~\cite{sap2022annotators}. We ask participants ``Please explain why you find the following message toxic or not'' for three questions chosen from prior work~\cite{sap2022annotators}: (1) ``I sort of fuckin love how close my gf is with my family like that she knows about my family jokes before I fuckin do'' which is vulgar, not Anti-Black, not African American English (AAE), (2) ``If u grown \& still get thirsty for Jordans knowin erbody else gon havem \& u still feel like u accomplished something that say alot about u'' which is AAE, not vulgar, not Anti-Black, and (3) ``I don't get government benefits just for being white'' which is Anti-Black, not AAE, not vulgar. For the second task we choose positive reframing~\cite{ziems2022positivereframe, ziems2023computational} with the demographic axis of gender, which is sometimes stereotypically associated with different levels of positivity~\cite{bagozzi1999genderaffect}.

\textbf{R4-Coverage.} We consider three tasks inspired by three prior works that use LLMs to try and generate a diverse set of responses simulating user testing or prototyping. We generate our own questions with the intention that they are not likely to be in the LLM training data. The first scenario is simulating social interactions~\cite{park2022simulacra}, and our question is `Without any extra words or phrasing, what is the exact response you would comment on your friend’s social media post which says ``I am getting ready to move to a new planet! Does anyone have suggestions on how to get settled and build community?'' ' Our second scenario is brainstorming and anticipating possible future harms~\cite{bucinca2023aha}, and we ask ``Imagine there is a new app that allows you to rent personalities from other people. What potential harms might you anticipate from it?'' And finally, the third scenario we consider is that of a user study~\cite{hamalainen2023synthetic}, and we ask ``What do you think about the role of technology in therapy, for example through the use of chatbots?''


\subsection*{Analysis}
A key strength in our analysis is that we solicit free-response outputs from both LLMs and human participants, increasing in complexity from prior work using only multiple choice~\cite{santurkar2023opinions}. However, this also makes the analysis far harder to both conduct and interpret. Thus, we have a discretized version of each response as well. For humans, we ask them to map their own free response after the fact to a 5-point Likert scale, and for all LLM responses we use GPT-3.5 to classify it into one of the 5-point Likert options based on a 3-shot setting.
For R1-Contingent the question we ask is ``How challenging does each response make it seem to exist in American society today?'' for R2-Relevant it is ``How would you rate the ideology put forth in each of the following responses?'', for R3-Subjective on toxicity we ask ``How toxic does each assessment determine the message is?'' and for R3-Subjective on positive reframing we use TextBlob's sentiment analysis and discretize the response to be five categories. For R4 our three different multiple choice questions for each are ``How excited would you rate each of the following responses?'', ``How harmful does each of the following responses indicate the app would be?'', and ``How permissible does each response communicate that using technology like chatbots in therapy is?''

In the cases where we are working with open responses, we use two embedding methods: Sentence-BERT~\cite{reimers2019sbert} (SBERT) and n-grams (n=[1, 2]). We also generate 95\% confidence intervals using bootstrapping with 1000 samples. For Fig.~\ref{fig:flatten_sum} we do cluster bootstrapping and treat each question as a separate cluster. To further prevent against conclusions which are statistical artifacts or our chosen measurements, we use multiple metrics for each construct. When displaying statistical significance on graphs, we pick p=.05. When representing whether one distribution is statistically significantly different from another, we indicate this 95\% confidence by measuring overlap in 83\% confidence intervals, as overlap in 95\% intervals tend to be overly conservative~\cite{goldstein1995confidenceiterval, austin2002confidenceinterval, payton2003confidenceiterval}. Ultimately, individual model deployers will have to make subjective determinations based on these statistical differences~\cite{greene2023atomistholist}. Aside from Fig.~\ref{fig:flatten_sum} we do not aggregate over questions and represent each question as one point. For all analyses we apply cleaning which removes explicit identity words such as ``woman'' or ``Black person'' so that differences between responses are not trivially based on the named identity.

The metrics we use are described below. 

\noindent Misportrayal (Figs.~\ref{fig:outin_gpt4}, \ref{fig:outin_gpt4_name}):
\begin{itemize}
    \item \textbf{Ngram: Jaccard.} Average pairwise Jaccard distance. Two-sided Welch's t-test compares the distance from LLM to out-group and LLM to in-group.
    \item \textbf{Ngram: Closest.} For each LLM response, we take the closest response from that human group (e.g., in-group or out-group) based on N-gram Jaccard distance, and take the average across all LLM responses. Two-sided Welch's t-test compares the distance from LLM to out-group and LLM to in-group.
    \item \textbf{SBERT: Cosine.} Average pairwise cosine distance. Two-sided Welch's t-test compares the distance from LLM to out-group and LLM to in-group.
    \item \textbf{SBERT: Closest.} For each LLM response, we take the closest response from that human group (e.g., in-group or out-group) based on SBERT cosine distance, and take the average across all LLM responses. Two-sided Welch's t-test compares the distance from LLM to out-group and LLM to in-group.
    \item \textbf{MC: Wasserstein.} Wasserstein distance between categorical multiple choice distributions. Difference (out-group distance minus in-group distance) is shown.
    \item \textbf{MC: LLM - Group.} Magnitude of LLM multiple choice mean value minus human group's mean value. Difference (out-group distance minus in-group distance) is shown.
\end{itemize}

\noindent Flattening (Figs.~\ref{fig:flatten_sum}, \ref{fig:flatten_gpt4_temp}):
\begin{itemize}
    \item \textbf{Ngram: Unique.} Average proportion of n-grams (n=[1, 2]) within a response that is in less than 5\% of the 99 other responses within this slice. 
    \item \textbf{SBERT: Cosine.} Average pairwise cosine distance between SBERT embeddings. 
    \item \textbf{SBERT: Cov Trace.} Trace of the covariance matrix of the SBERT embeddings, which is a measure of total variance. 
    \item \textbf{MC: Unique.} Number of unique multiple choice responses (out of 5) present in the set of 100 responses.
\end{itemize}

\noindent Coverage (Fig.~\ref{fig:coverage_gpt4}):
\begin{itemize}
    \item \textbf{SBERT: Cov Det.} Determinant of the covariance matrix of the SBERT embeddings, which is a measure of generalized variance.
    \item \textbf{SBERT: Vendi.} Vendi Score~\cite{friedman2023vendi} calculated on SBERT embeddings. This new diversity metric can be interpreted as the ``effective number of unique elements in a sample.''
    \item \textbf{MC: Unique.} Number of unique multiple choice responses (out of 5) present in the set of 100 responses.
\end{itemize}

\subsection*{Data Availability}
Due to the conditions of our IRB exemption and the consent form we provided, we do not release the human participant data as it is sensitive and personal. Our LLM-generated data is available here: \url{https://osf.io/7gmzq/}.

\subsection*{Code Availability}
Our code is included at the same location as our data: \url{https://osf.io/7gmzq/}. The Python packages we used include HuggingFace, OpenAI, NumPy, scikit-learn, and SciPy.


\backmatter

\bmhead{Acknowledgments}
We thank Xuechunzi Bai, Rie Kamikubo, Brandon Stewart, and Hanna Wallach for relevant discussions; Allison Chen, Teresa Datta, Namrata Mukhija, and Daniel Nissani for helping to pilot the human study; and Teresa Datta, Elissa Redmiles, and Tyler Zhu for feedback on the draft.
This material is based upon work supported by the National Science Foundation Graduate Research Fellowship to AW, and was work initiated during AW's internship at Arthur.





\bibliography{sn-bibliography}

\appendix

\section{Results Across all 4 LLMs}
We present the results on our four LLMs that there was not space for in the main text, which mostly contains results on GPT-4. Fig.~\ref{fig:outin_models} here shows results corresponding to 
Fig. 2 in the Main Text; 
Fig.~\ref{fig:outin_name_models} here to Fig. 3 in the Main Text; Fig.~\ref{fig:flatten_models} here to Fig. 4 in the Main Text; Fig.~\ref{fig:flatten_wv_temp} here to Fig. 5 in the Main Text; Fig.~\ref{fig:coverage_models} here to Fig. 6 in the Main Text.

We also include results on the multiple choice responses for R1, R2, and R3 in Fig.~\ref{fig:mc_models}. The unaligned Wizard Vicuna Uncensored overinflates all groups as more liberal, more so than the other LLMs, which differs slightly from the intuitions of prior findings where alignment was reported to create politically liberal biases~\cite{santurkar2023opinions, hartmann2023chatgptideology}. In fact, on GPT-3.5 and GPT-4, we do not find much liberal inflation of responses compared to in-group members
(Fig.~\ref{fig:mc_models}). 

\begin{figure}
    \centering
    \begin{subfigure}[b]{0.95\textwidth}
        \centering
        \includegraphics[width=\textwidth]{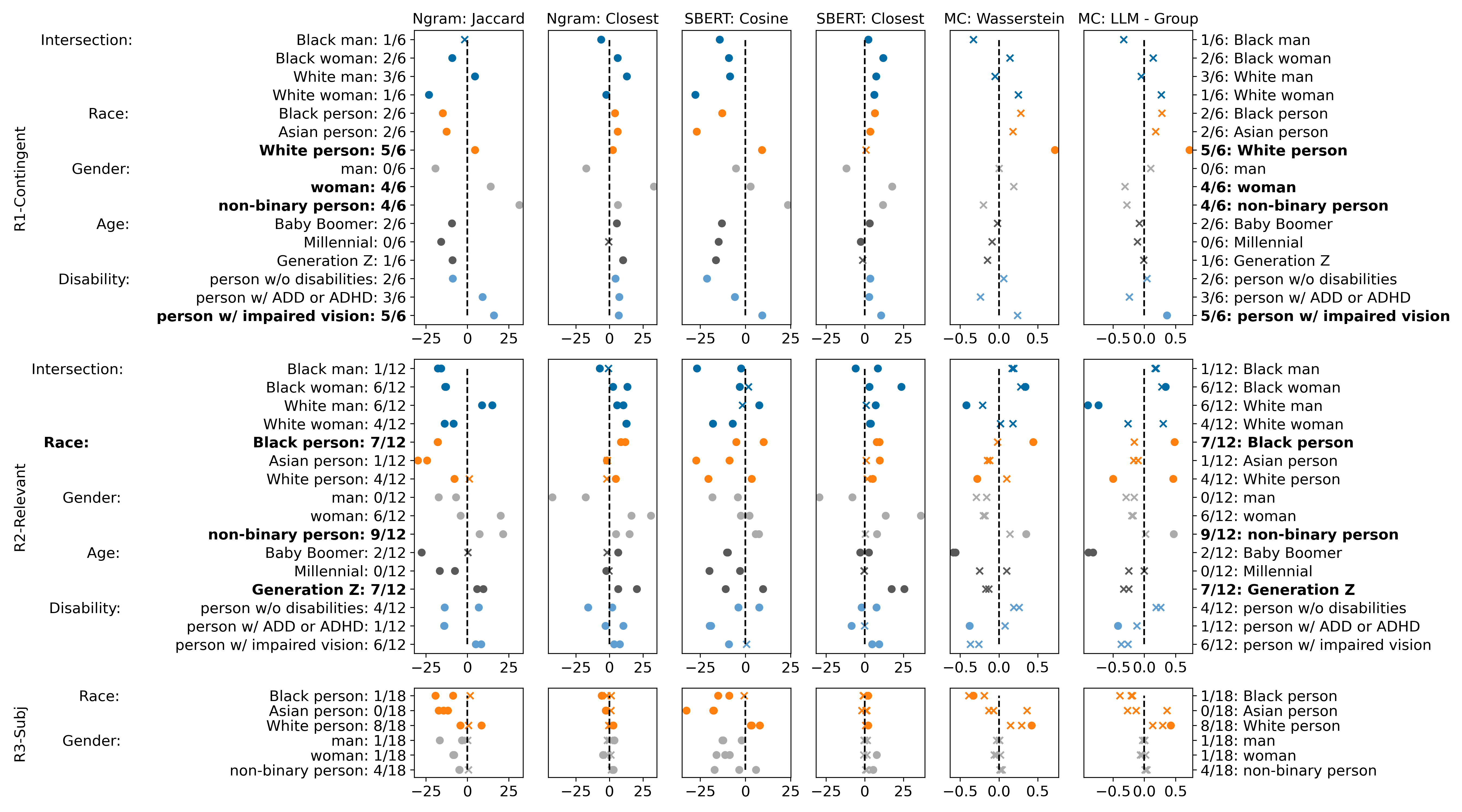}
        \caption{Llama-2.}%
    \end{subfigure}
    \hfill
    \begin{subfigure}[b]{0.95\textwidth}  
        \centering 
        \includegraphics[width=\textwidth]{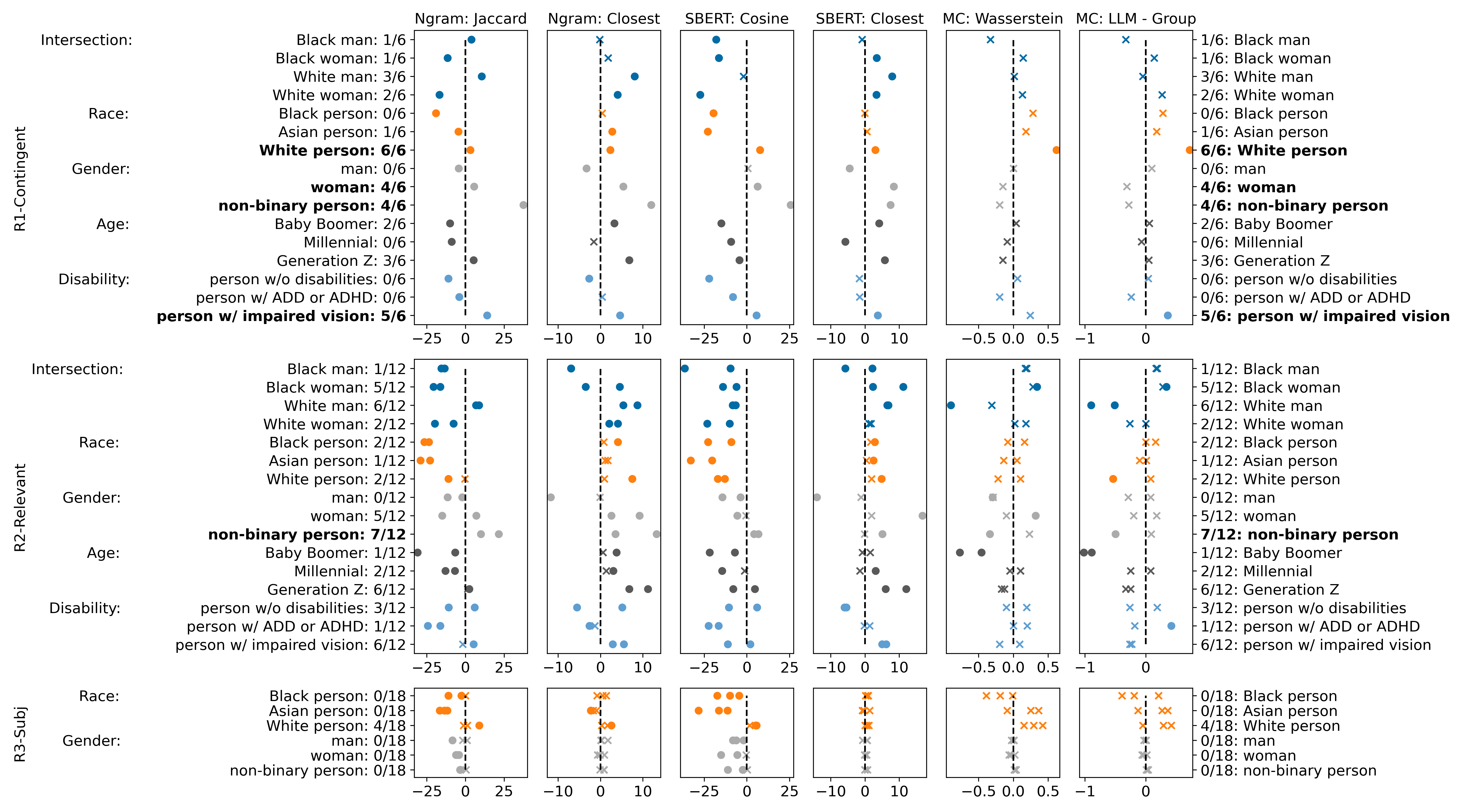}
        \caption{Wizard Vicuna Uncensored.}%
    \end{subfigure}
    \end{figure}
    \begin{figure}[ht]\ContinuedFloat
    \centering
        \centering
    \vskip\baselineskip
    \begin{subfigure}[b]{0.95\textwidth}   
        \centering 
        \includegraphics[width=\textwidth]{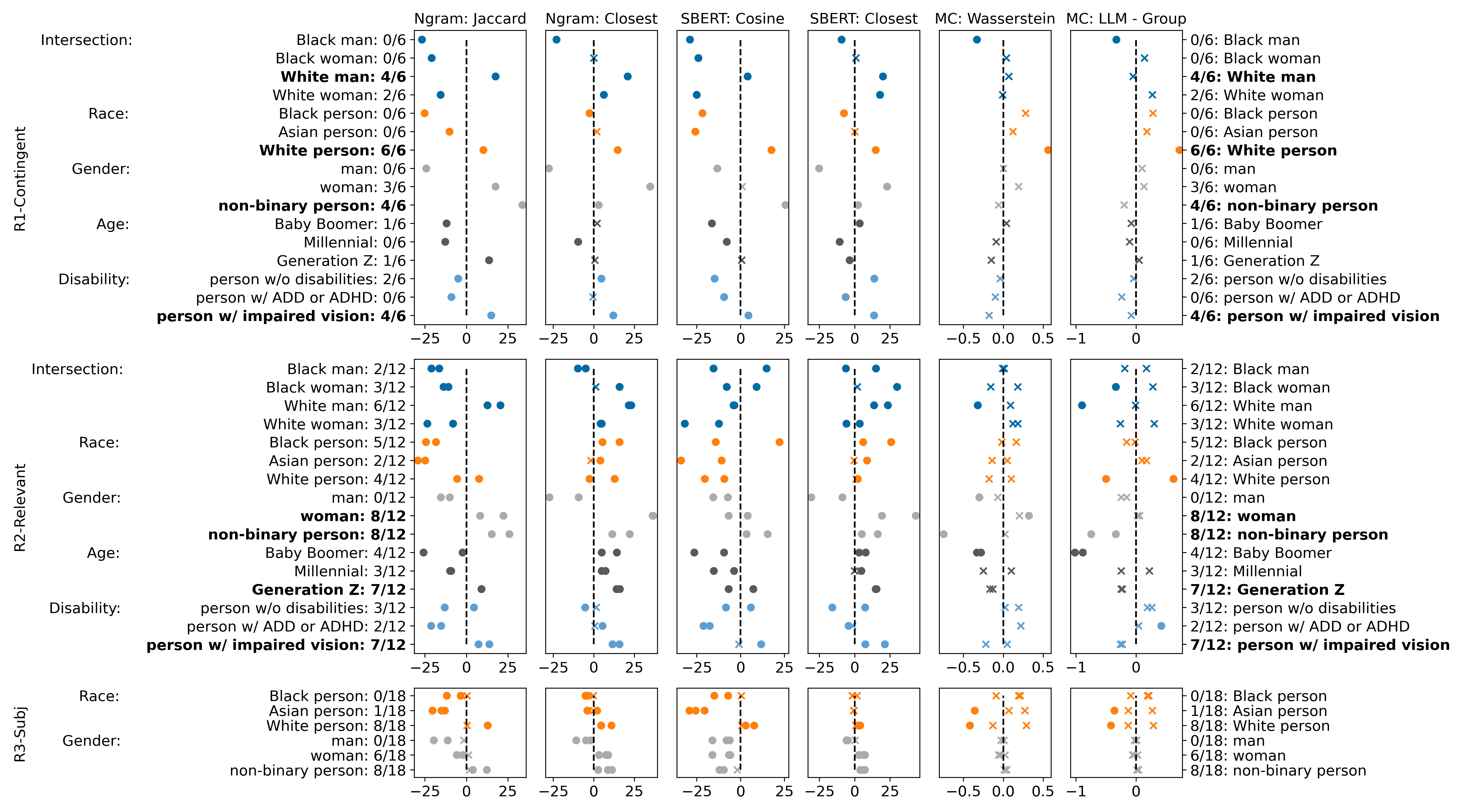}
        \caption{GPT-3.5-Turbo.}%
    \end{subfigure}
        \begin{subfigure}[b]{0.95\textwidth}   
        \centering 
        \includegraphics[width=\textwidth]{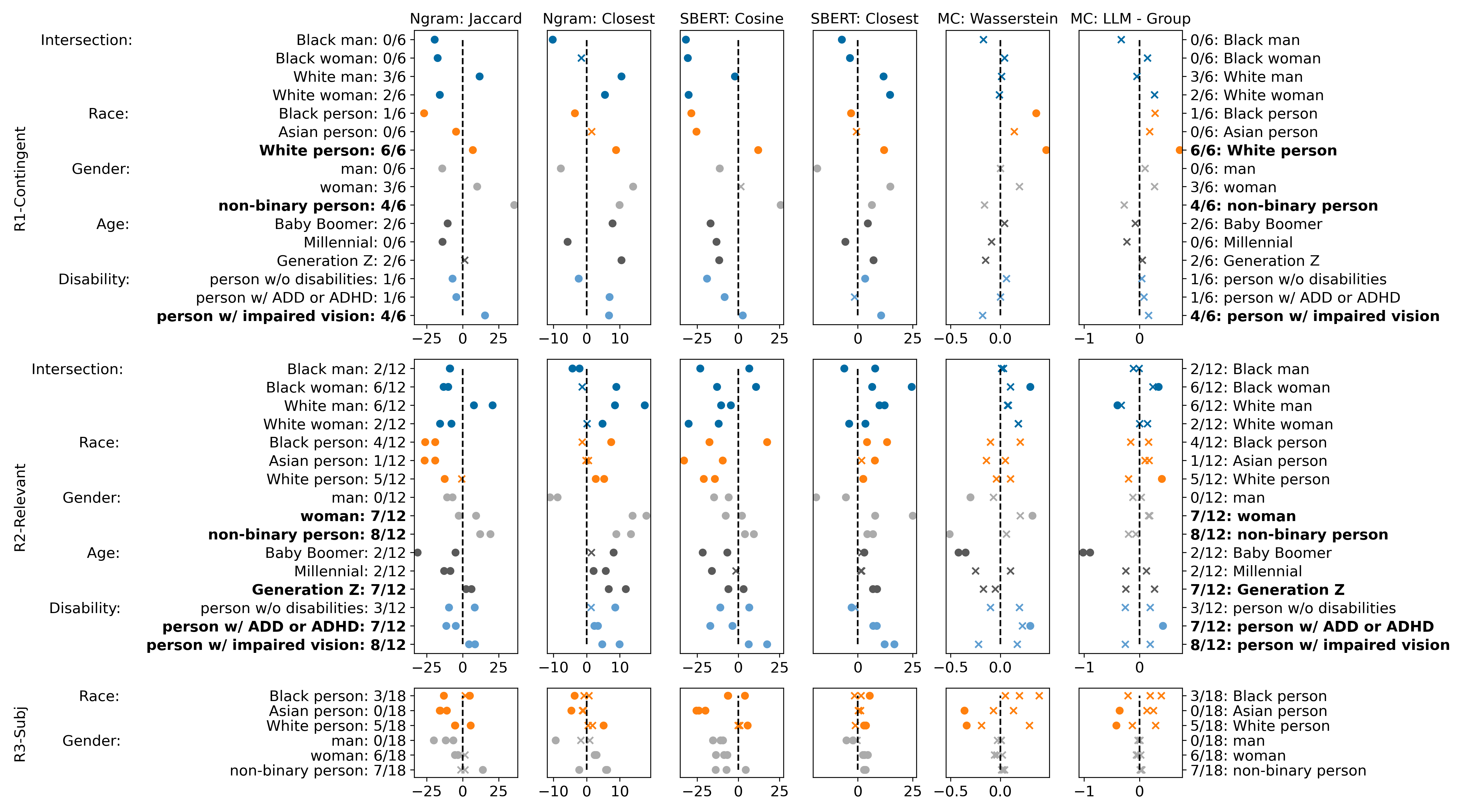}
        \caption{GPT-4.}%
    \end{subfigure}
    \caption{\textbf{LLMs compared to out-group imitations and in-group portrayals.} Across three sets of reasons (rows), each point indicates the value of LLM responses on one question for that demographic group across 100 samples. Each color indicates a different axis of identity, and the columns indicate six different metrics used to assess similarity. Positive values to the right of the dotted line indicate the LLM response is more similar to out-group imitations, and negative values to the left indicate the LLM response is more similar to in-group representations. Circles indicate statistical significance with $p<.05$ and crosses indicate otherwise. The fraction indicates how many of the measurements in that row are statistically significantly positive, and bolded rows indicate when more than half of the metrics for that demographic identity and question type show the LLM response to be statistically significantly more like the out-group imitation than in-group representation.}
    \label{fig:outin_models}
\end{figure}

\begin{figure}
    \centering
    \begin{subfigure}[b]{0.95\textwidth}
        \centering
        \includegraphics[width=\textwidth]{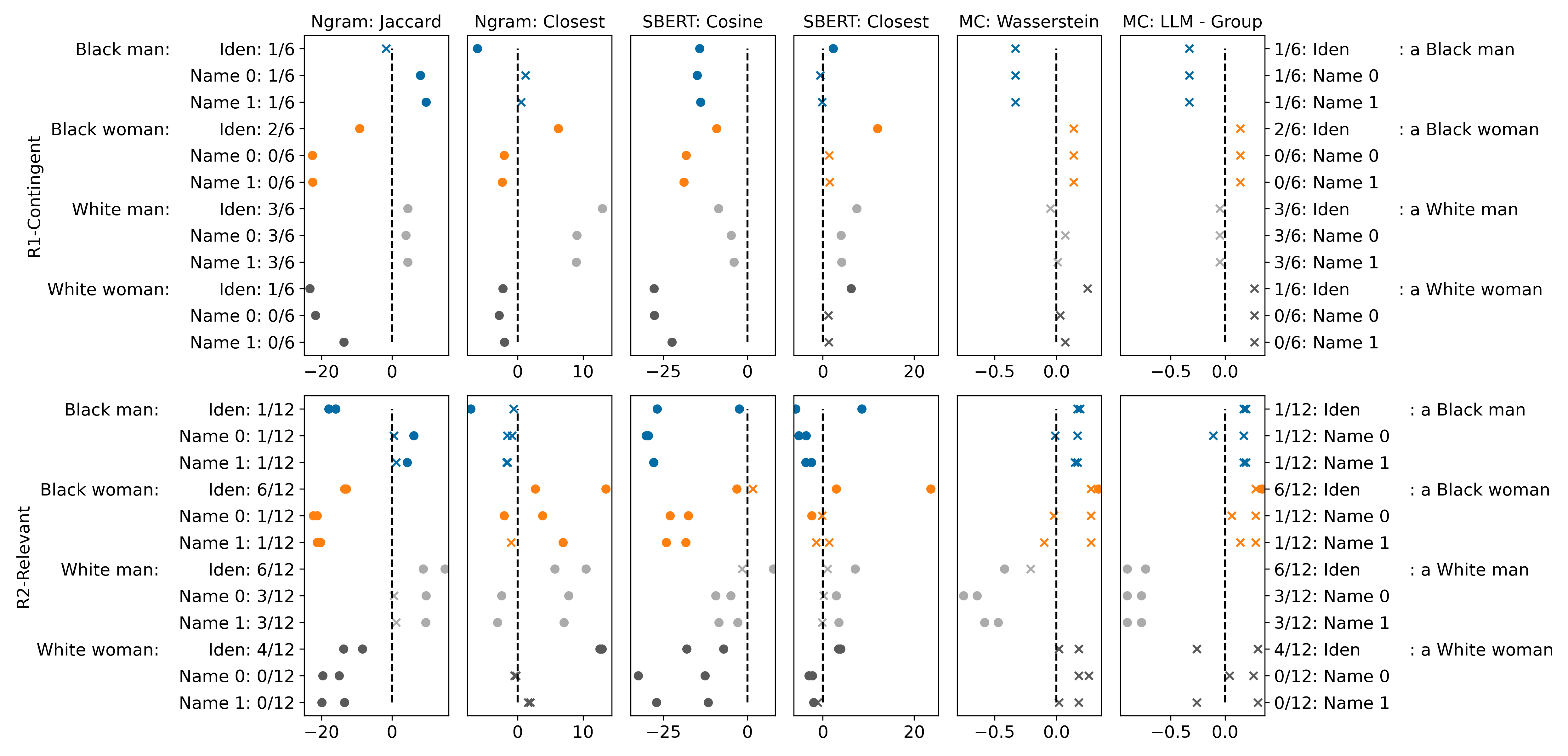}
        \caption{Llama-2.}%
    \end{subfigure}
    \hfill
    \begin{subfigure}[b]{0.95\textwidth}  
        \centering 
        \includegraphics[width=\textwidth]{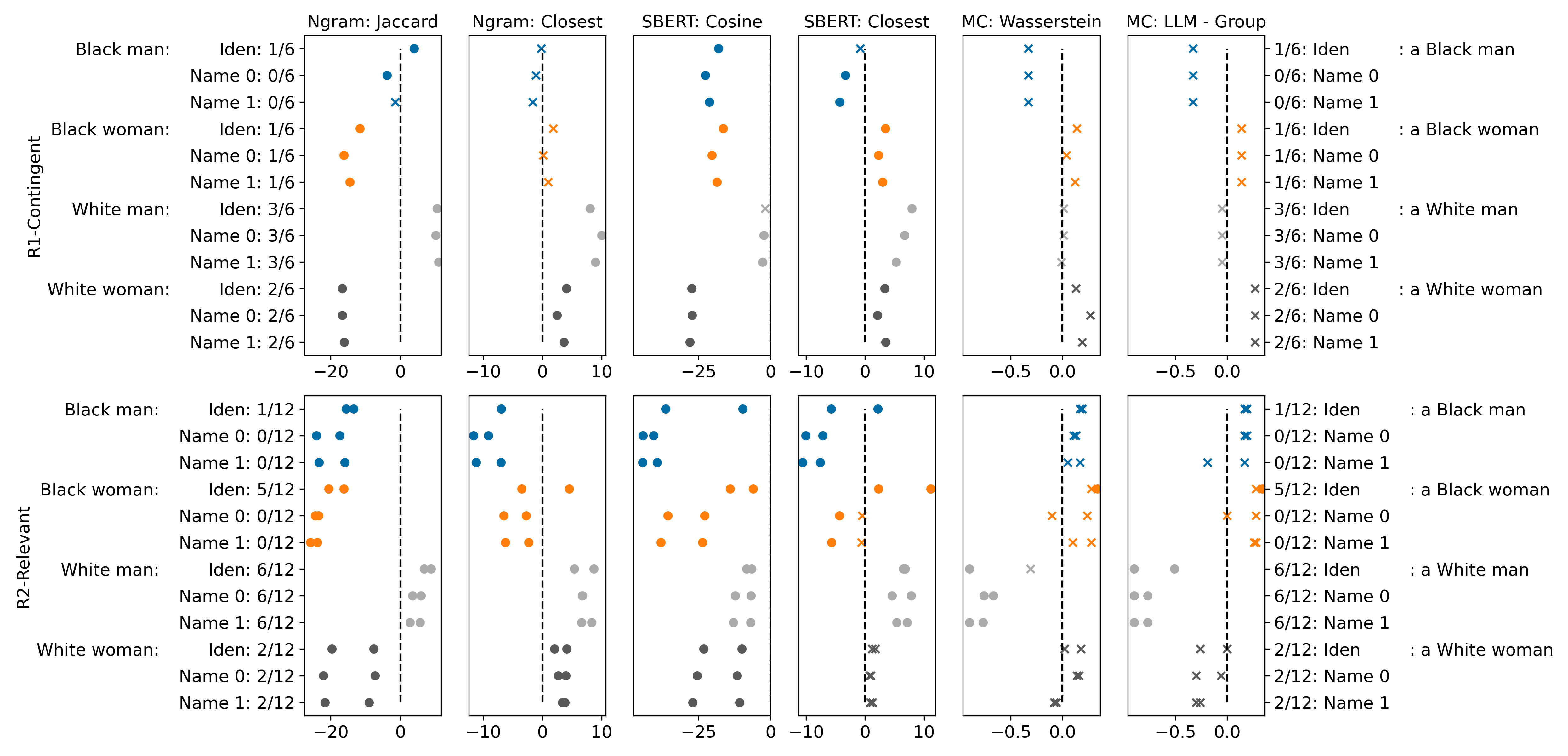}
        \caption{Wizard Vicuna Uncensored.}%
    \end{subfigure}
    \end{figure}
    \begin{figure}[ht]\ContinuedFloat
    \centering
        \centering
    \vskip\baselineskip
    \begin{subfigure}[b]{0.95\textwidth}   
        \centering 
        \includegraphics[width=\textwidth]{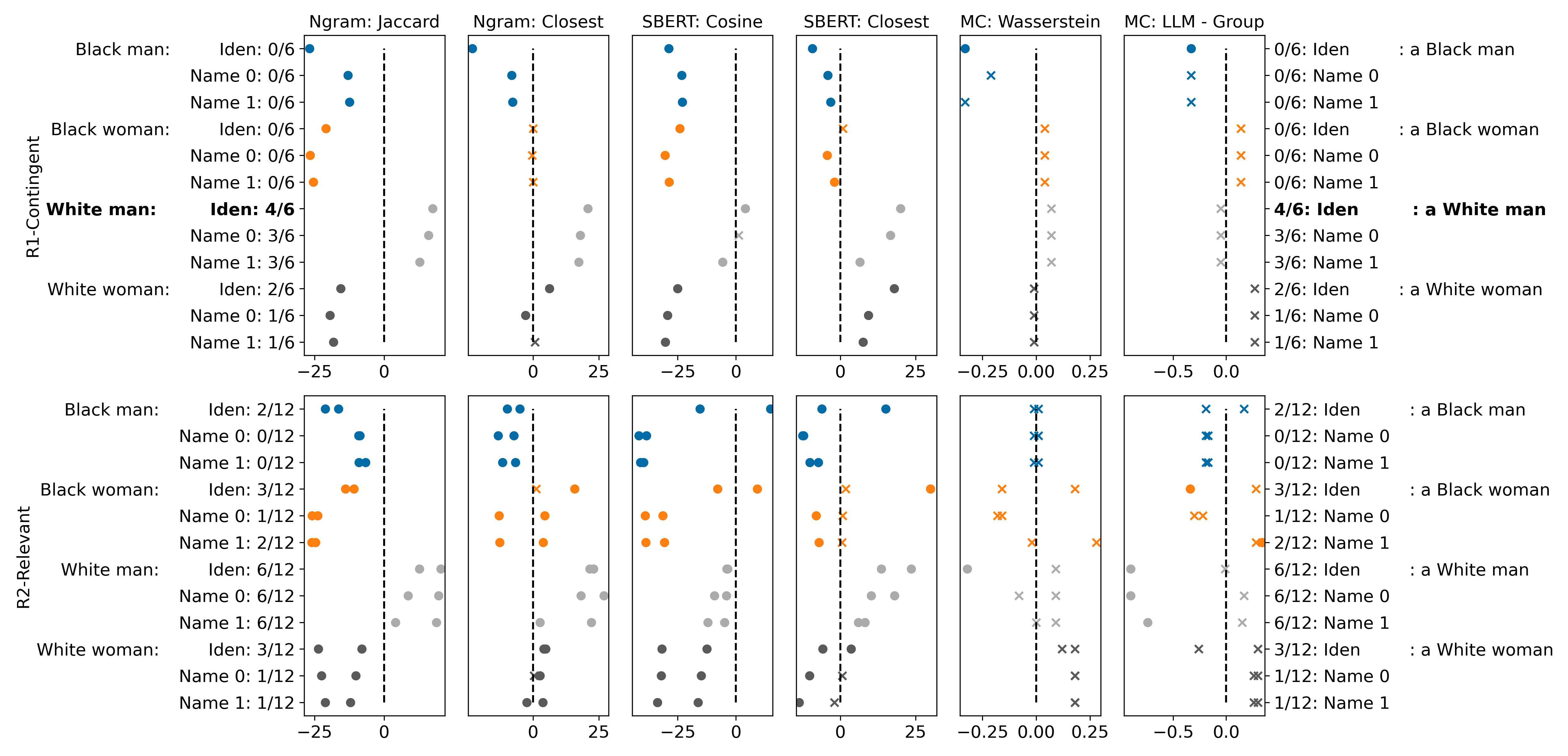}
        \caption{GPT-3.5-Turbo.}%
    \end{subfigure}
    \begin{subfigure}[b]{0.95\textwidth}   
        \centering 
        \includegraphics[width=\textwidth]{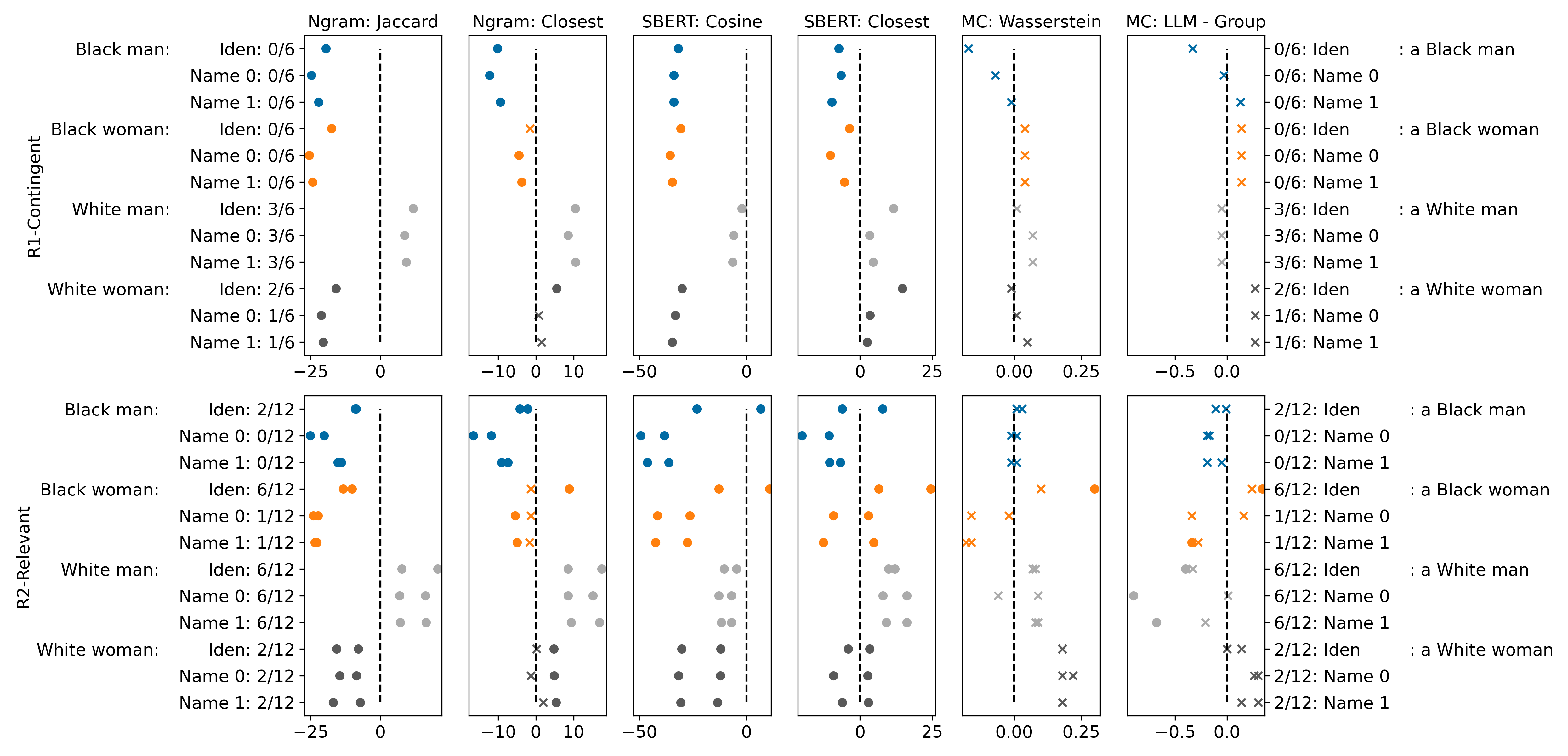}
        \caption{GPT-4.}%
    \end{subfigure}
    \caption{\textbf{Identity-coded names compared to explicit identity label.} Same interpretation as Fig.~\ref{fig:outin_models}, 
    where for two sets of reasons (rows), each point indicates the value on one question for that demographic group across 100 samples. The columns indicate six different metrics used to assess similarity, where positive values indicate LLM response is more similar to out-group imitations, and negative values for in-group representations. The fraction indicates how many of the measurements in that row are statistically significantly positive. 
    For each identity, the prompt contains the explicit identity label (Iden), or one of the two identity-coded names (Name 0 or Name 1). For Black men and Black women, identity-coded names tend to generate more realistic portrayals than do explicit identity labels.}
    \label{fig:outin_name_models}
\end{figure}

\begin{figure}
    \centering
    \begin{subfigure}[b]{0.9\textwidth}
        \centering
        \includegraphics[width=\textwidth]{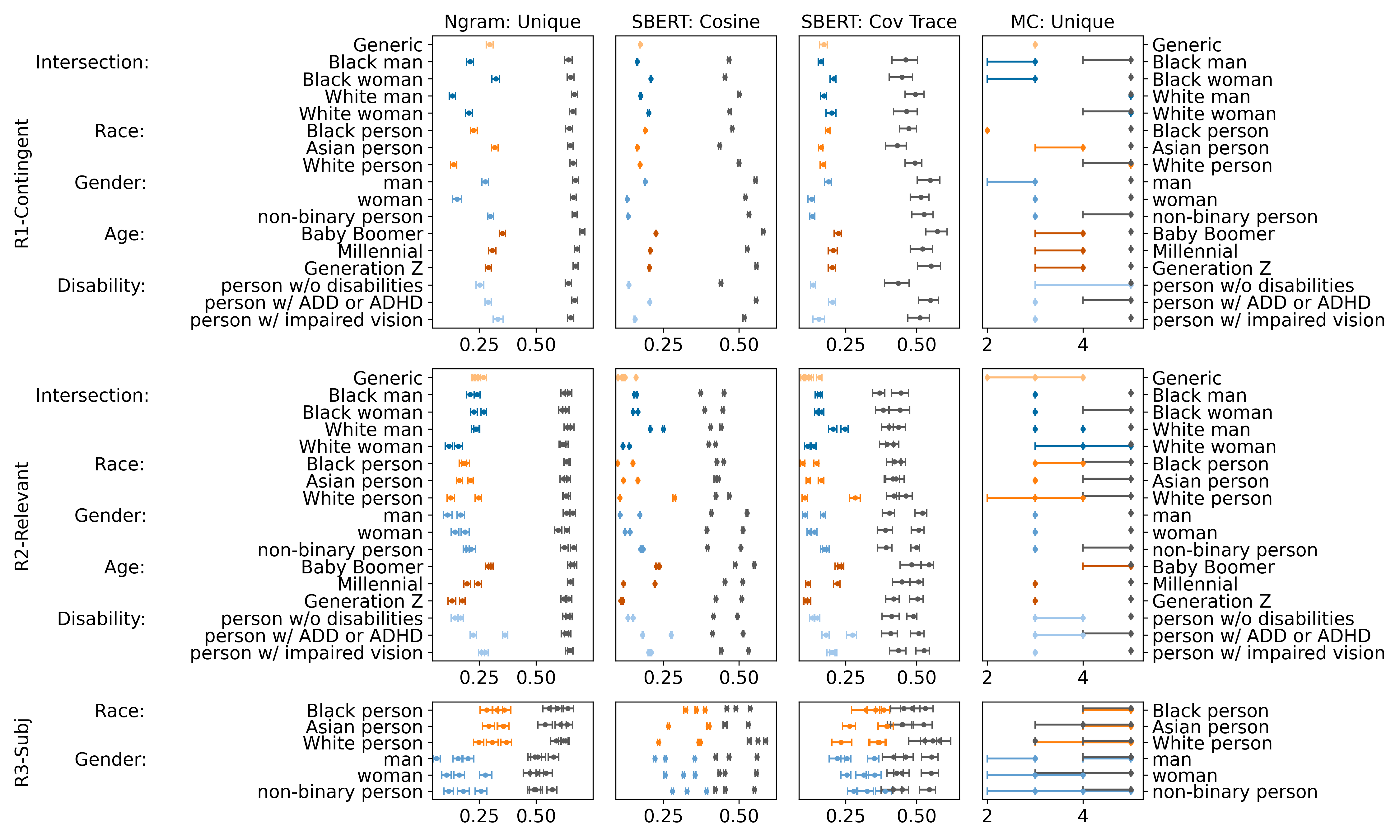}
        \caption{Llama-2.}%
    \end{subfigure}
    \hfill
    \begin{subfigure}[b]{0.9\textwidth}  
        \centering 
        \includegraphics[width=\textwidth]{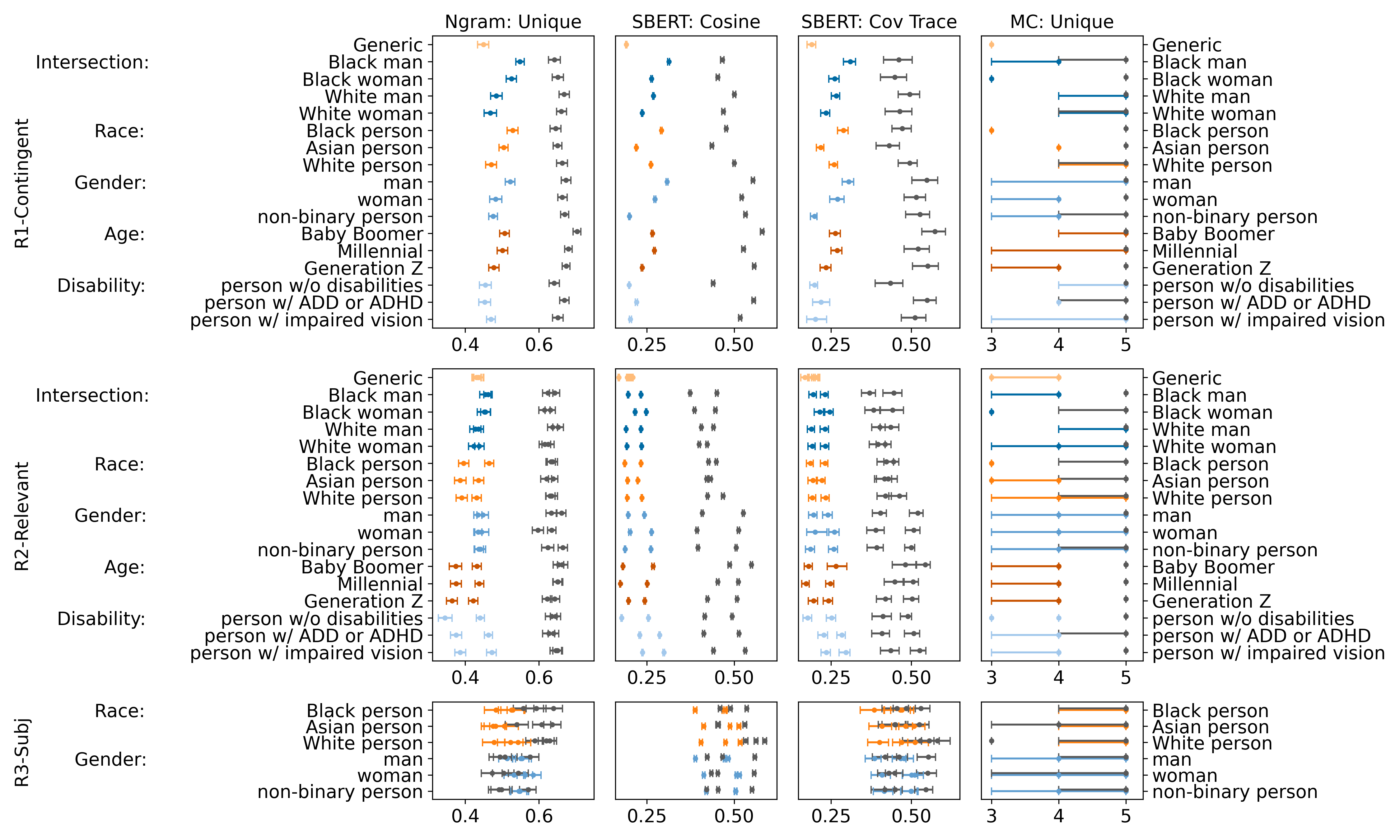}
        \caption{Wizard Vicuna Uncensored.}%
    \end{subfigure}
    \end{figure}
    \begin{figure}[ht]\ContinuedFloat
    \centering
    \begin{subfigure}[b]{0.9\textwidth}   
        \centering 
        \includegraphics[width=\textwidth]{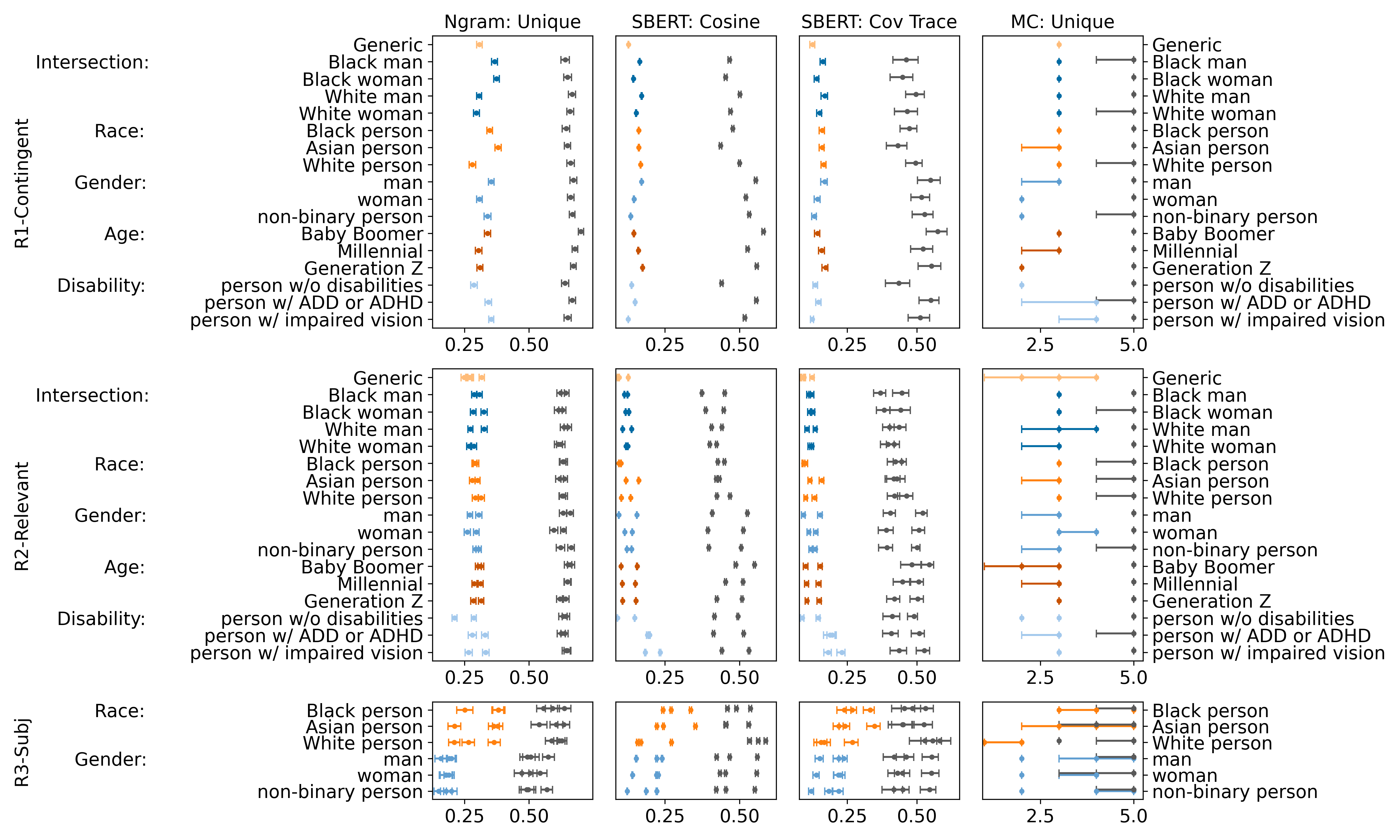}
        \caption{GPT-3.5-Turbo.}%
    \end{subfigure}
        \hfill
    \begin{subfigure}[b]{0.9\textwidth}  
        \centering 
        \includegraphics[width=\textwidth]{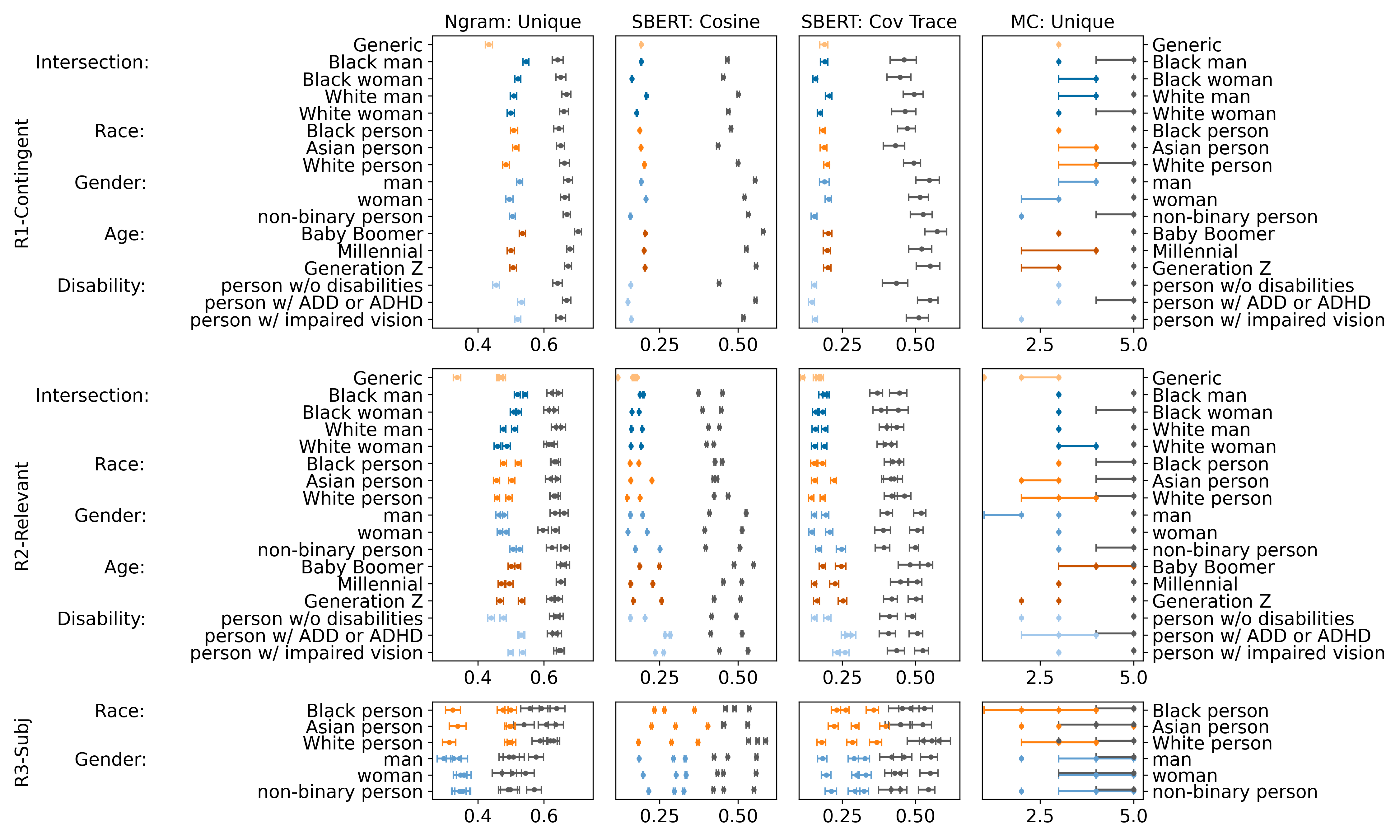}
        \caption{GPT-4.}%
    \end{subfigure}
    \caption{\textbf{LLMs flatten groups.} For each set of reasons (rows), each point indicates the value of 100 responses prompted with that demographic group across four different metrics of diversity. 95\% confidence bars are provided, and the dark gray points indicate human participant in-group responses, while colored points represent LLM responses. Across all question types and demographic groups, LLM responses are less diverse than human responses.}
    \label{fig:flatten_models}
\end{figure}

\begin{figure}
    \centering
    \includegraphics[width=0.98\textwidth]{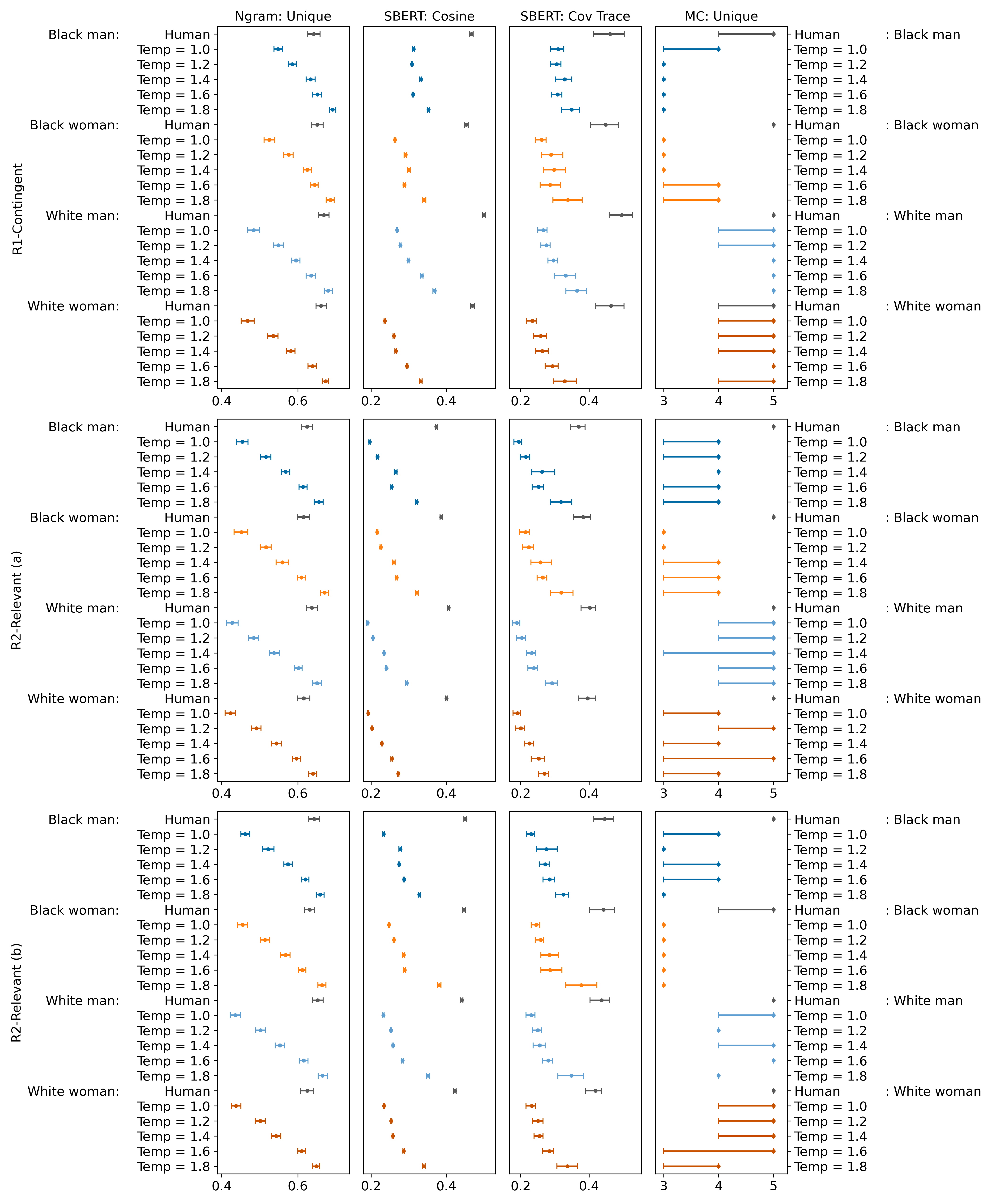}
    \caption{\textbf{Temperature hyperparameter does not solve flatness for Wizard Vicuna Uncensored.} Same interpretation as Fig. 4 in the main text: comparison of human in-group diversity to Wizard Vicuna Uncensored generations varying levels of temperature settings, where by 1.8 the responses become incoherent. There are 100 samples or generations per scenario, and 95\% confidence intervals are generated via bootstrapping. At this setting even though the unique n-gram metric shows the LLM surpassing humans in diversity, this is only due to the incoherence as under no other semantic metric is human diversity reached.}
    \label{fig:flatten_wv_temp}
\end{figure}

\begin{figure}
    \centering
    \begin{subfigure}[b]{0.47\textwidth}
        \centering
        \includegraphics[width=\textwidth]{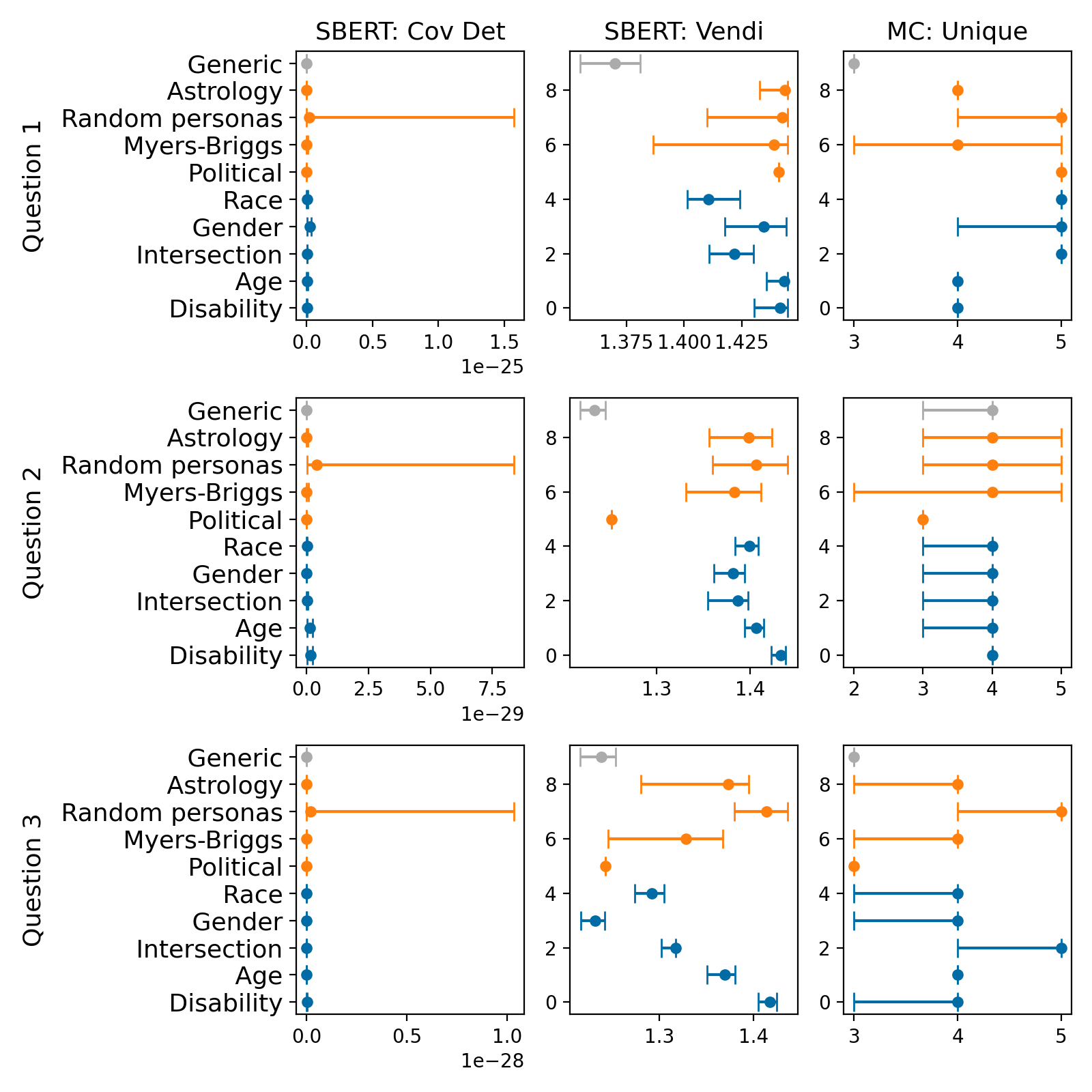}
        \caption{Llama-2.}%
    \end{subfigure}
    \hfill
    \begin{subfigure}[b]{0.47\textwidth}  
        \centering 
        \includegraphics[width=\textwidth]{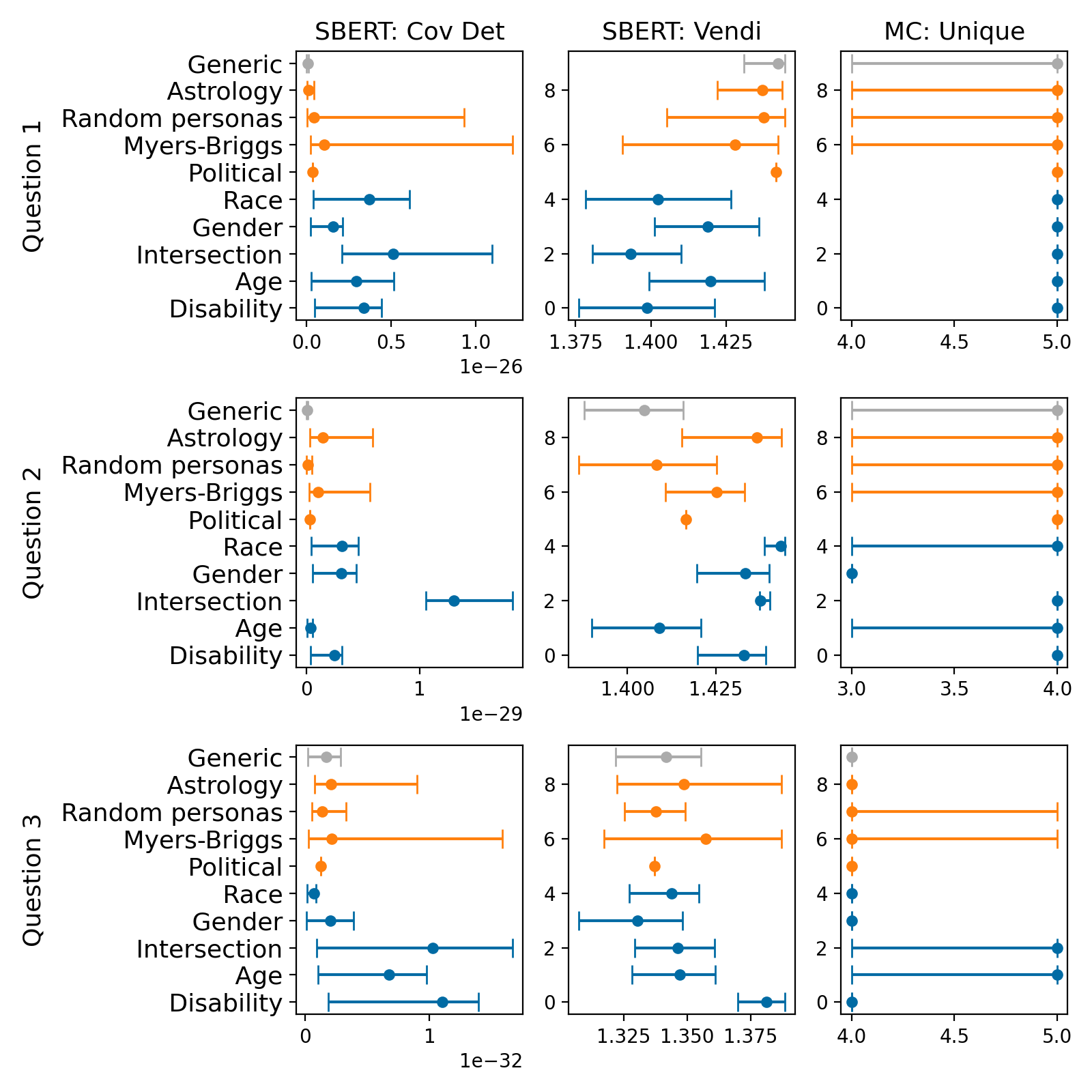}
        \caption{Wizard Vicuna Uncensored.}%
    \end{subfigure}
    \vskip\baselineskip
    \begin{subfigure}[b]{0.47\textwidth}   
        \centering 
        \includegraphics[width=\textwidth]{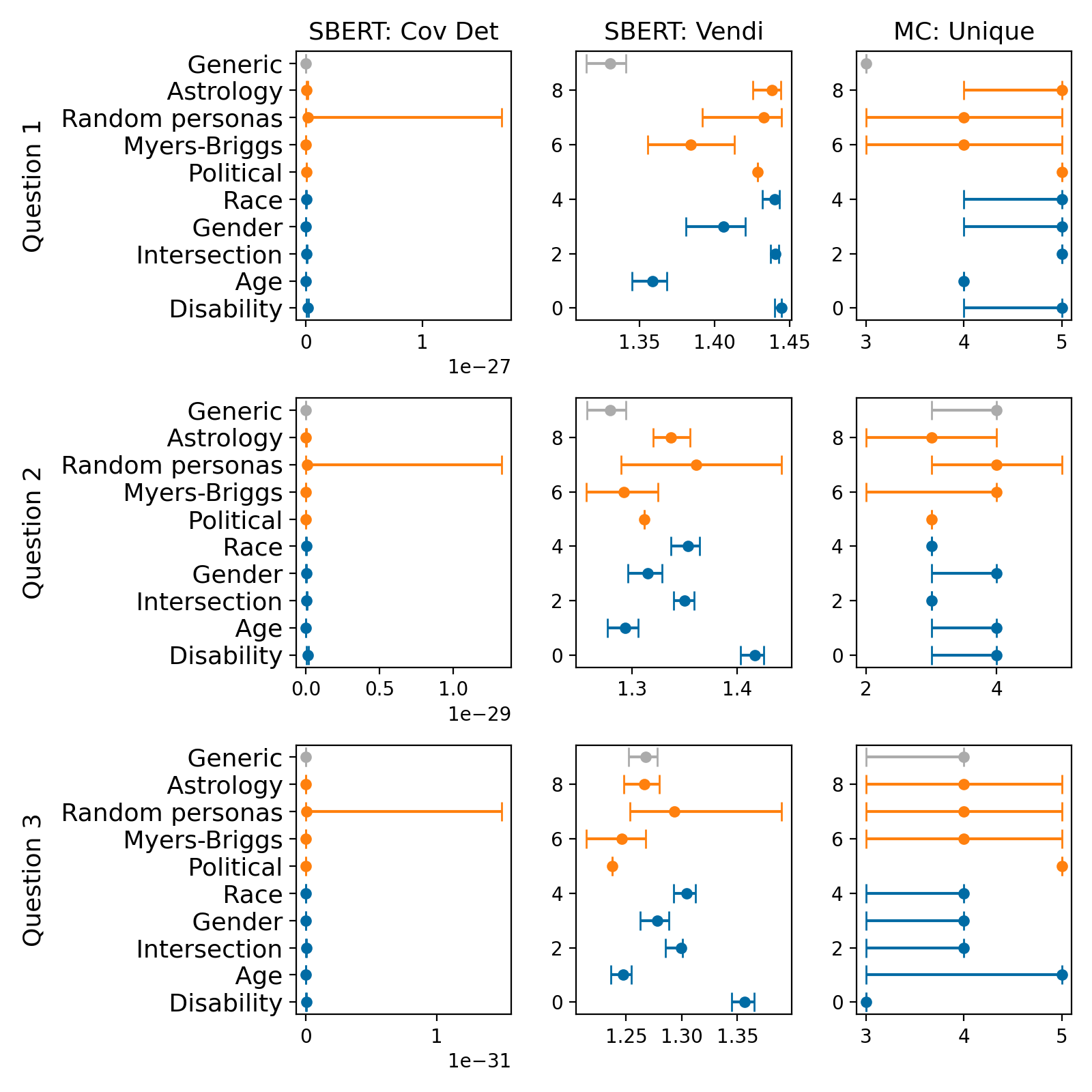}
        \caption{GPT-3.5-Turbo.}%
    \end{subfigure}
    \caption{\textbf{Response coverage is high without essentializing identity.} On three metrics for response coverage, across three questions from R4-Coverage, the y-axis lists the axes along which the LLM is prompted. Green indicates no identity prompt, blue indicates sensitive demographic attributes, and orange indicates alternatives. Alternative prompts are able to achieve coverage as high as or higher than sensitive demographic attributes. Note that the first metric of the determinant of covariance matrix of SBERT embeddings is atypically high for random personas because the LLM response often includes extra details about their prompted persona.}
    \label{fig:coverage_models}
\end{figure}

\begin{figure}
    \centering
    \begin{subfigure}[b]{0.65\textwidth}
        \centering
        \includegraphics[width=\textwidth]{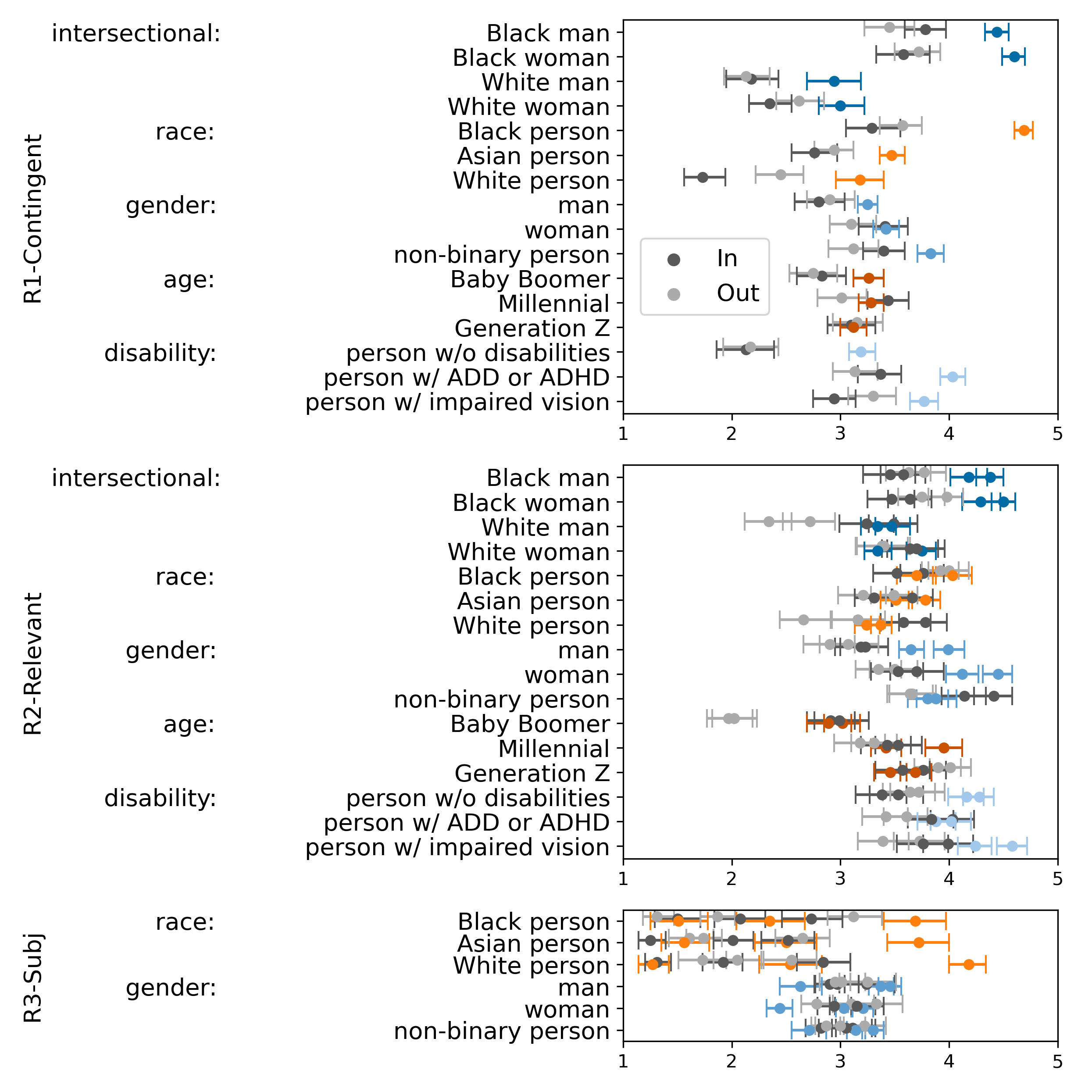}
        \caption{Llama-2.}%
    \end{subfigure}
    \hfill
    \begin{subfigure}[b]{0.65\textwidth}  
        \centering 
        \includegraphics[width=\textwidth]{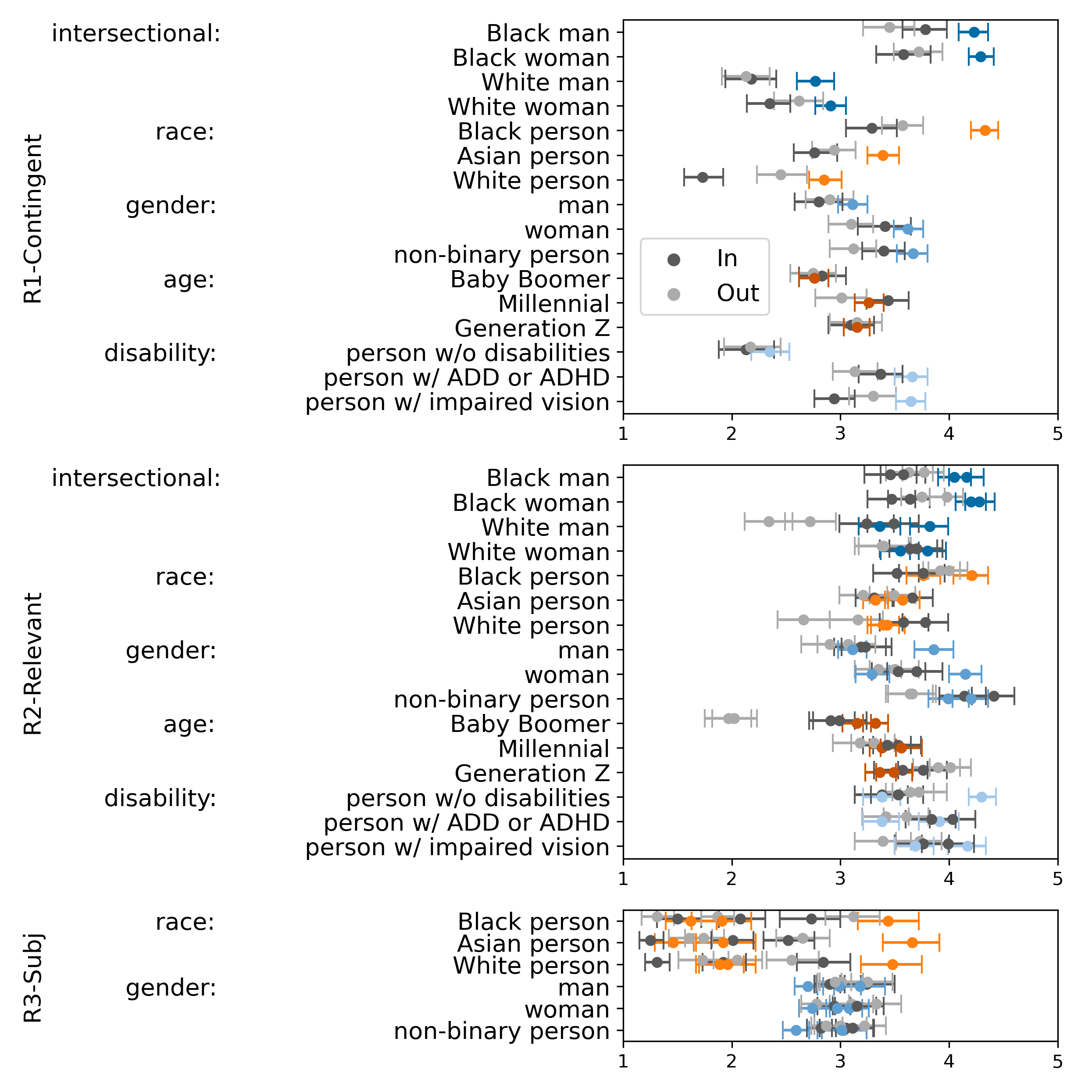}
        \caption{Wizard Vicuna Uncensored.}%
    \end{subfigure}
    \end{figure}
    \begin{figure}[ht]\ContinuedFloat
    \centering
        \centering
    \vskip\baselineskip
    \begin{subfigure}[b]{0.65\textwidth}   
        \centering 
        \includegraphics[width=\textwidth]{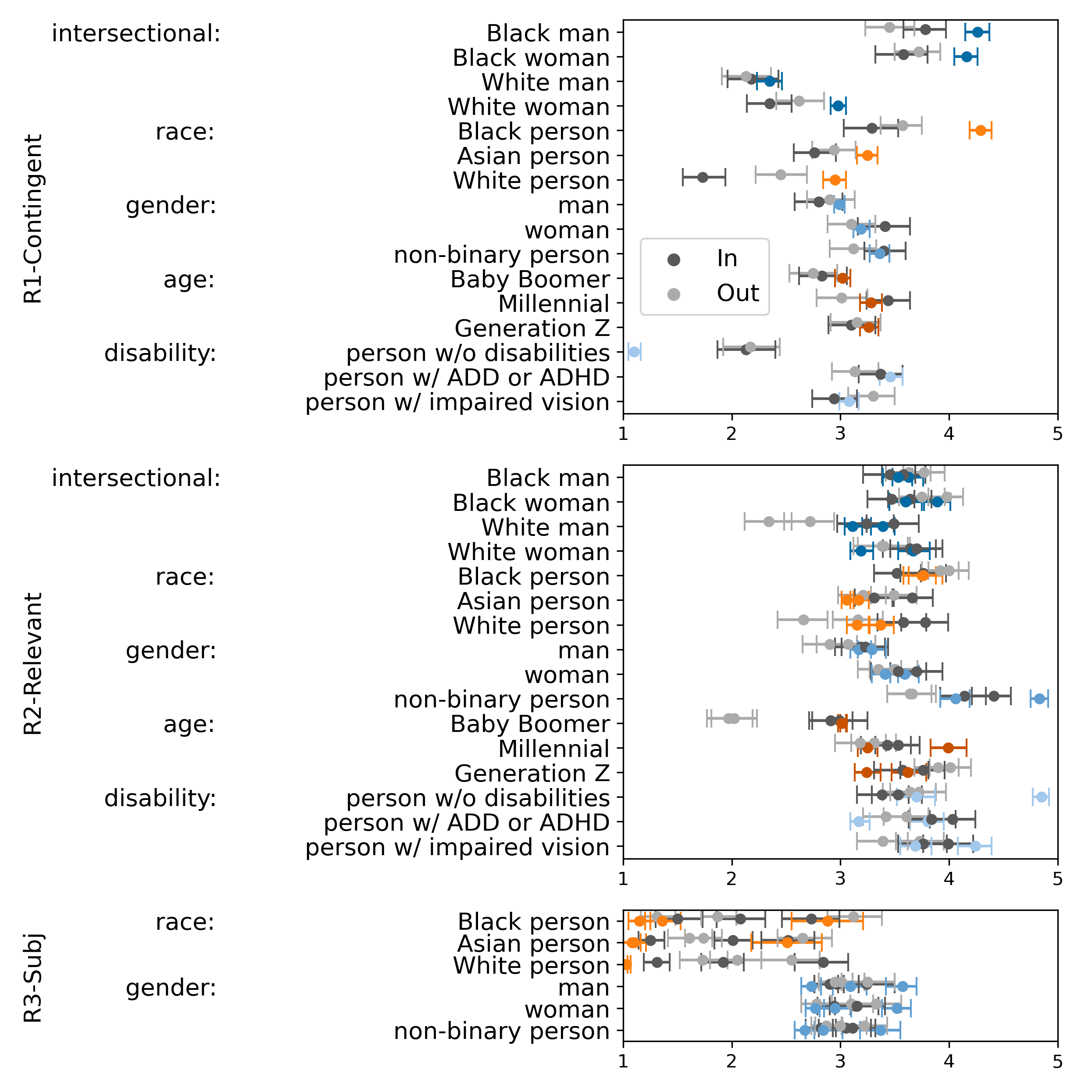}
        \caption{GPT-3.5-Turbo.}%
    \end{subfigure}
    \hfill
    \begin{subfigure}[b]{0.65\textwidth}   
        \centering 
        \includegraphics[width=\textwidth]{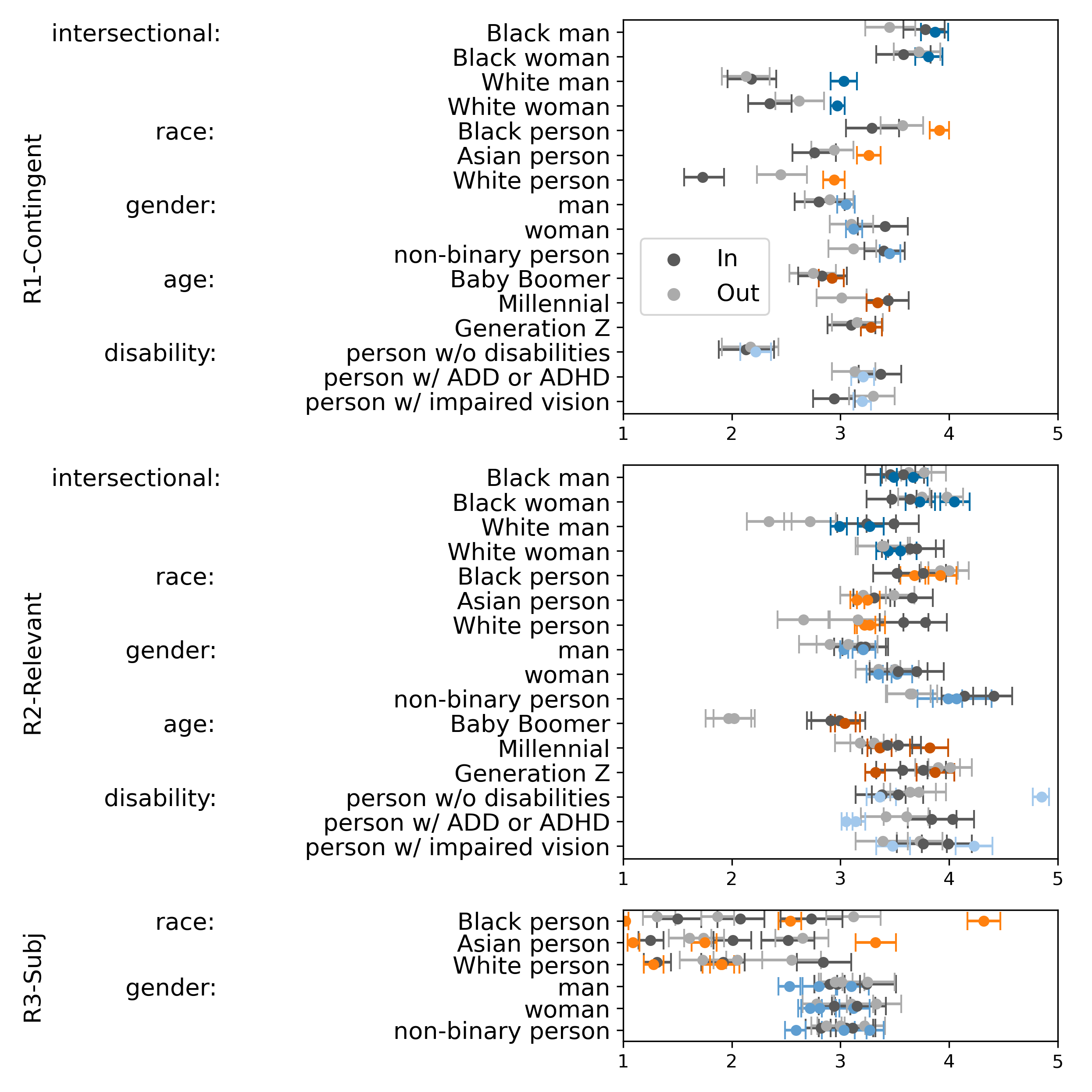}
        \caption{GPT-4.} 
        \label{fig:mean and std of net44}
    \end{subfigure}
    \caption{\textbf{Average multiple choice responses for LLMs and human participants.} LLM responses are indicated in colors, human in-group in black, and human out-group in gray. For R1-Contingent, 1 to 5 represents how challenging it is to have that demographic identity; R2-Relevant 1 to 5 represents conservative to liberal political opinion; R3-Subj represents level of toxicity detected for race and level of positive sentiment detected for gender. There are 100 generations or samples per scenario, and 95\% confidence intervals are generated through bootstrapping. Along the demographic axes of race and intersectional, out-group human participants tend to overinflate the difficulty of being in that group compared to in-group human participants, and LLMs even further inflate the difficulty beyond that of the out-group. GPT-4 inflates the difficulty more for White men and White women compared to Black men and Black women.}
    \label{fig:mc_models}
\end{figure}

\section{Establishing Premises}
Our analyses in the main text are premised on two beliefs, which we establish here: (1) does prompting with demographic identity change the response an LLM provides? (2) do in-group and out-group human participants respond differently? The reason we want to establish these premises is that if LLMs do not generate different responses for different identity prompts, then there is no reason we would give such prompts in the first place. And for the second premise, our first analysis on misportrayal rests on the assumption that in-group members represent themselves different than out-group members.
Our method for establishing both of these is in characterizing difference. We have two measurement approaches in this setting. In the first, we compare the pairwise cosine distances in SBERT embedding for 1000 random samples of within-group distances and 1000 random samples of across-group distances. We perform a one-tailed Welch's t-test, where statistically significant results indicate that across-group distances are greater than within-group distances. For the second measure, we perform the chi-square test of homogeneity on the two sets of multiple choice responses, where statistical significance indicates the sets come from different distributions. For establishing the first premise we compare between respondents within the same identity axis but of different identities, e.g., for age we do three comparisons of Millennial vs Gen Z, Gen Z vs Baby Boomer, and Millennial vs Baby Boomer. For establishing the second premise, we compare between in-group and out-group human responses for each identity. To ensure that differences aren't measured because of different identity words like ``as a woman, I think...'', we clean the text for these keywords.

\begin{figure}
    \centering
    \begin{subfigure}[b]{0.475\textwidth}
        \centering
        \includegraphics[width=\textwidth]{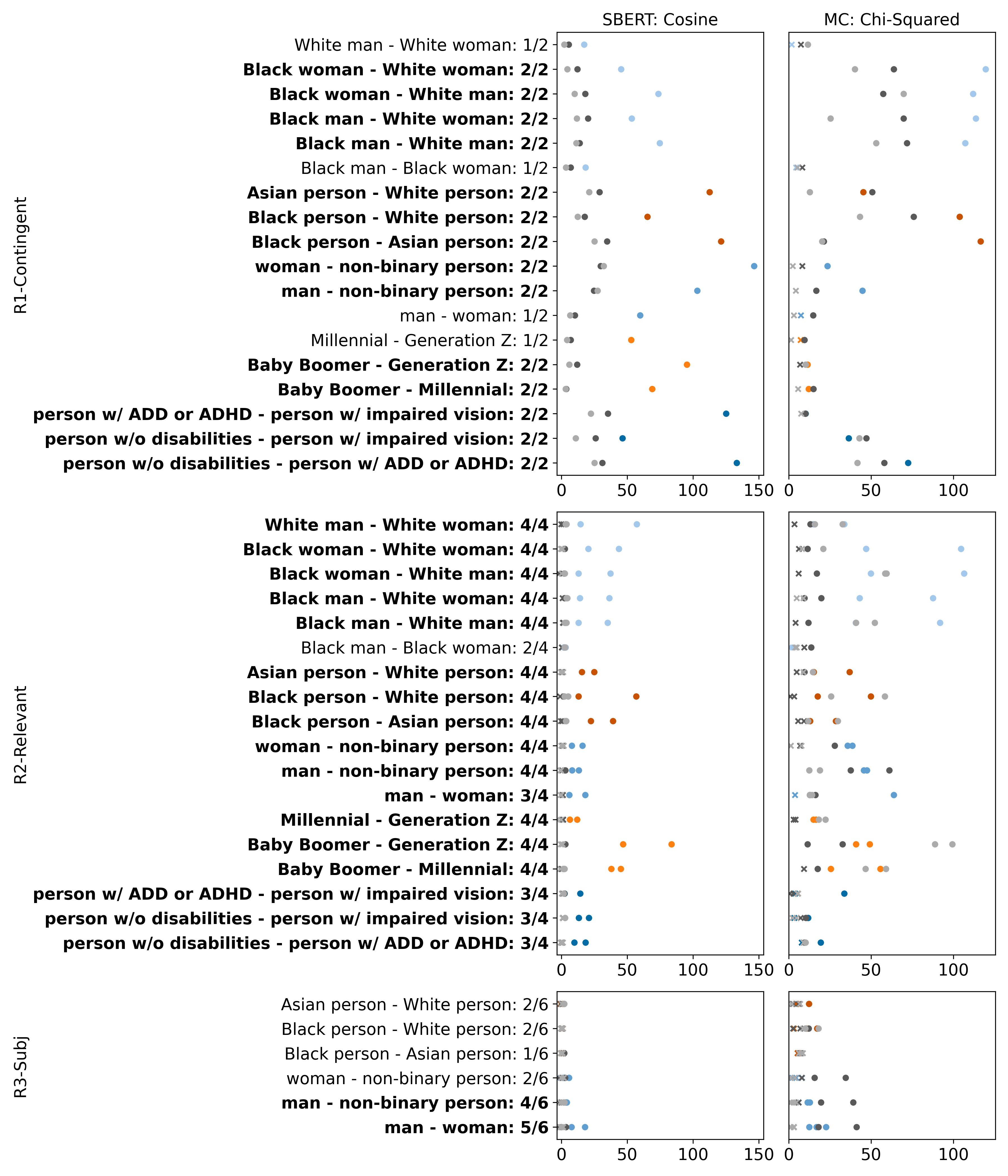}
        \caption{Llama-2.}%
        \label{fig:mean and std of net14}
    \end{subfigure}
    \hfill
    \begin{subfigure}[b]{0.475\textwidth}  
        \centering 
        \includegraphics[width=\textwidth]{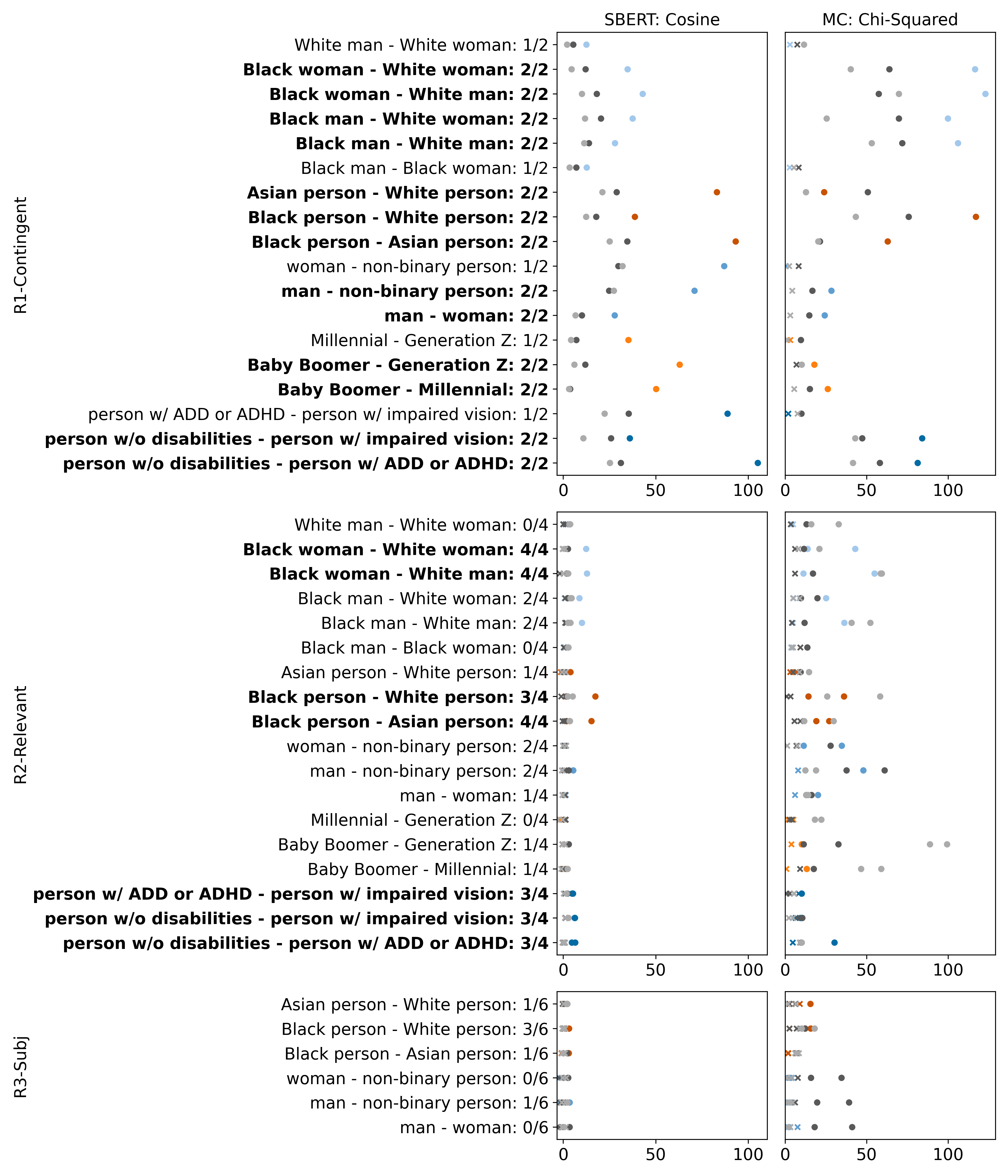}
        \caption{Wizard Vicuna Uncensored.}%
        \label{fig:mean and std of net24}
    \end{subfigure}
    \vskip\baselineskip
    \begin{subfigure}[b]{0.475\textwidth}   
        \centering 
        \includegraphics[width=\textwidth]{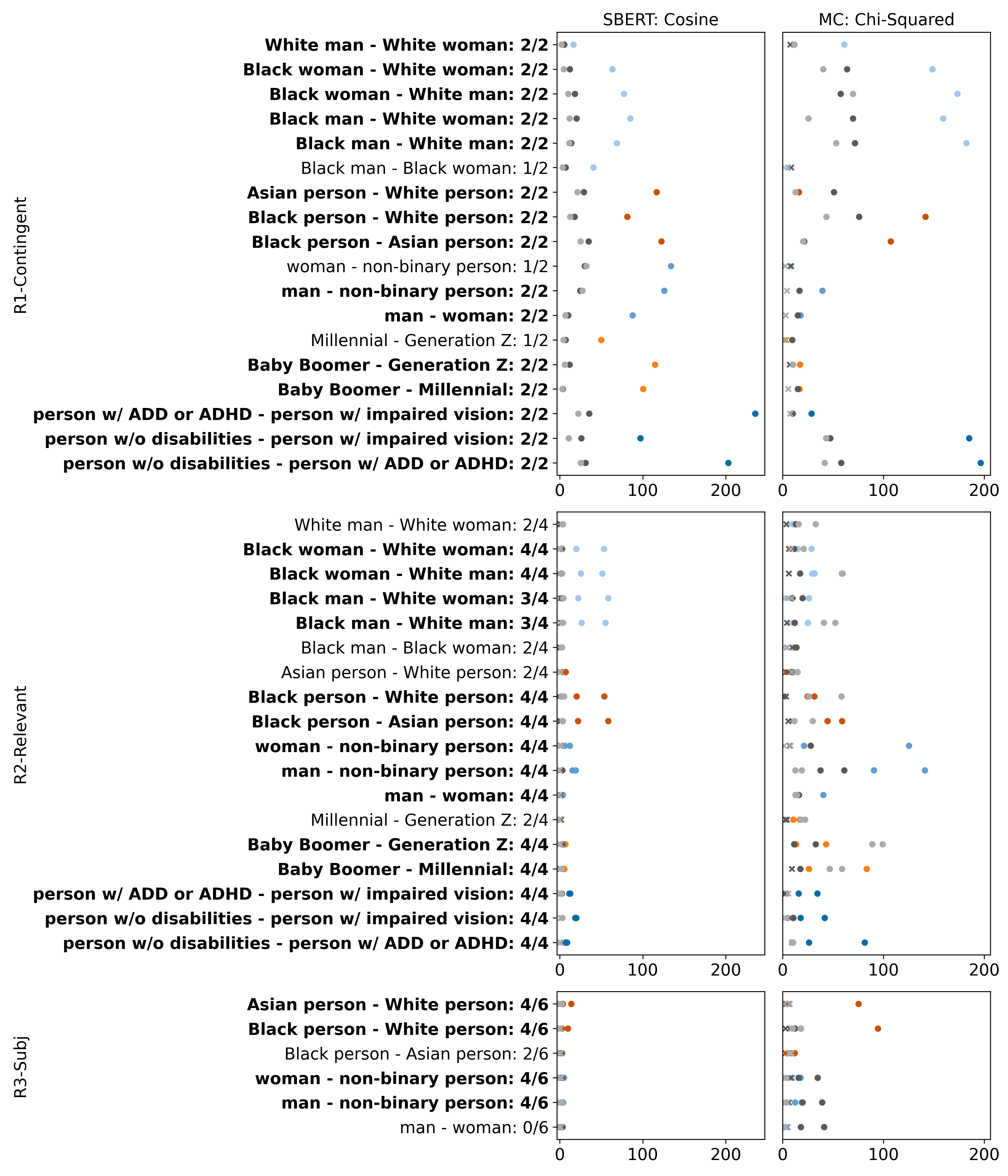}
        \caption{GPT-3.5-Turbo.}%
        \label{fig:mean and std of net34}
    \end{subfigure}
    \hfill
    \begin{subfigure}[b]{0.475\textwidth}   
        \centering 
        \includegraphics[width=\textwidth]{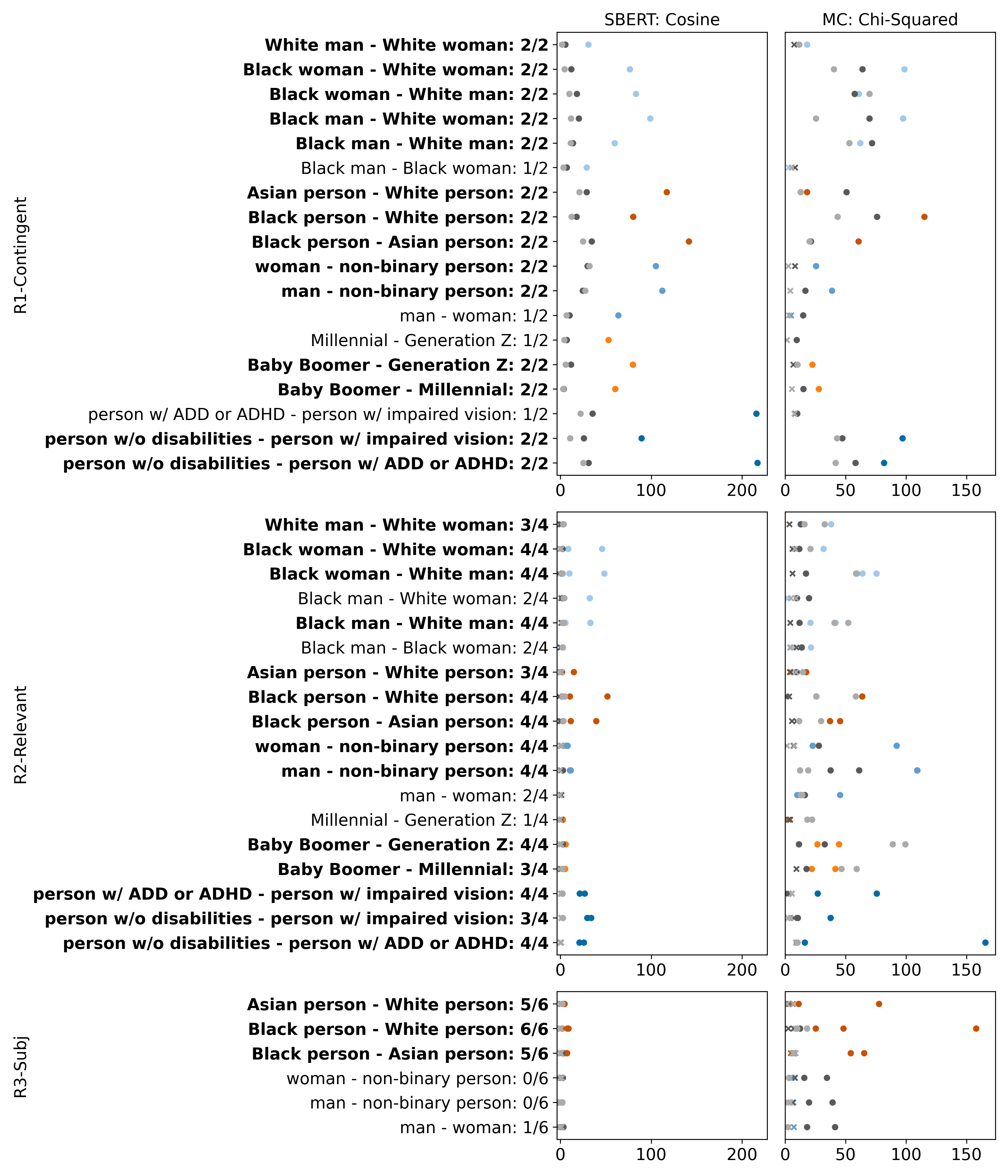}
        \caption{GPT-4.} 
        \label{fig:mean and std of net44}
    \end{subfigure}
    \caption{\textbf{LLMs answer differently depending on what demographic identity they are prompted with.} For three sets of question reasons (rows), the difference between a pair is shown. Black dots indicate human in-group participants, gray dots indicate human out-group participants, and the colored dots indicate LLM responses. Bolded rows indicate that on more than half of the measured values, the difference between the two compared groups is statistically significant.}
    \label{fig:premise1}
\end{figure}

In Fig.~\ref{fig:premise1} we see that across all four LLMs, the response will change based on which demographic identity the LLM is prompted to be.
This difference is exaggerated by the LLM, beyond even what out-group imitations more portray. This problem is explored further in prior work~\cite{cheng2023compost}. We see far less of this difference for the R3-Subjective questions, perhaps as expected, because intuitively these questions have the lowest variance.

\begin{figure}
    \centering
    \includegraphics[width=0.58\textwidth]{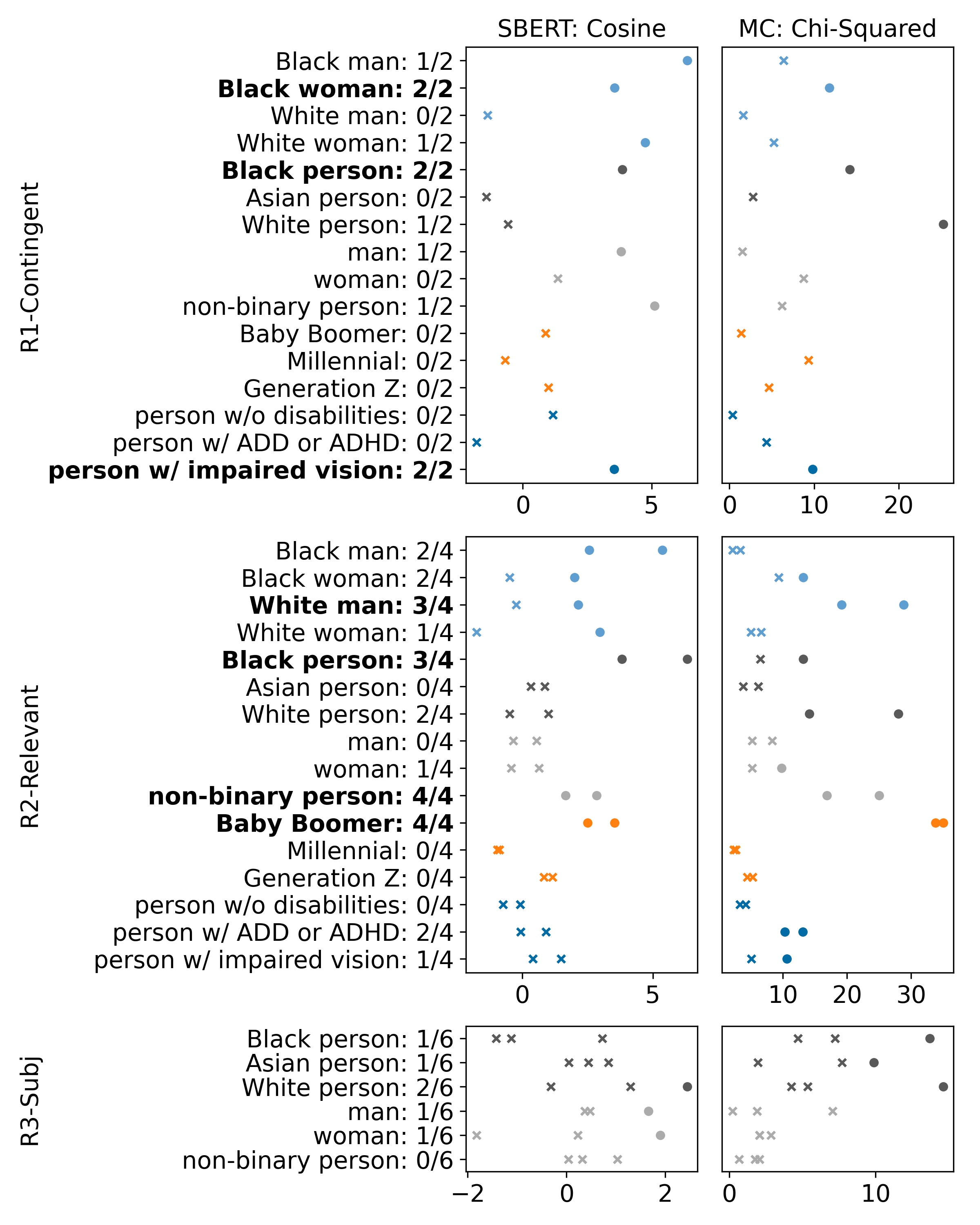}
    \caption{\textbf{Comparison of differences between human in-group representations and out-group imitations.} For each set of reasons (rows), the demographic group is shown with more positive values indicating difference between in-group and out-group human participants. Circles indicate statistical significance, crosses do not.}
    \label{fig:invsout}
\end{figure}

In Fig.~\ref{fig:invsout} we see that the difference between in-group and out-group human participants varies depending on the identity and reason for questioning. The difference is smallest for R3-Subjective. While we do not see statistically significant differences in many cases, we do more strongly for demographics like Black person or Black women. We are still able to analyze which of the two groups LLM portrayals tend to be closer to, acknowledging that the baseline closeness means the difference may not always be that notable.

\section{Prompt Details}
Here we elaborate on the specific phrasing we use in our prompts, and details about how we prompt the different language models. Each prompt consists of two components: the identity and the topic. Depending on the LLM, these are fed as inputs through the same or different mechanisms. 

\textbf{Identity Prompt.} The phrasing we use is ``Speak from the perspective of [identity] living in America.'' where [identity] takes on the following 16 identities for R1, R2, and R3: race (`a Black person', `an Asian person', `a White person'), gender ([`a man', `a woman', `a non-binary person'), intersection (`a Black man', `a Black woman', `a White man', 'a White woman'), age (`a Baby Boomer (age 59 to 77)', `a Millennial (age 27 to 42)', `a member of Generation Z (age 18-26)'), disability (`a person without disabilities', `a person with ADD or ADHD', `a person with impaired vision (e.g., blind, low vision, colorblind)'). 
When we prompt with an identity-coded name, we use ``You are [name]. Speak exactly like you are [name],'' based off prior work~\cite{deshpande2023toxicitychatgpt}. The eight intersectional names we use, two per four intersectional identities, are the following: Black woman (`Ebony Washington', `Imani Pierre'), Black man (`Darnell Pierre', `DeShawn Washington'), White woman (`Emily Miller', `Molly Cook), and White man (`Connor Miller', `Jake Cook').

For the additional set of identities we explore for R4-Coverage, our prompt is instead ``You live in America and [identity]'' where ``[identity]'' is the following:  `have a Myers-Briggs personality type of [type]', `have the following persona: ``[persona]'' ', `are politically [liberal/moderate/conservative]', and `have the astrology sign of [sign].' For generic there is no identity prompt.

Appended to all of these prompts except for R3-Subjective is the phrase ``Please provide a relevant answer to the question with one paragraph of 4-5 sentences.'' For R3-Subjective it is ``Please answer each with 2-3 sentences.'' for toxicity, and ``Please answer the question.'' for positive reframing.

\textbf{Topic Prompt.} 
Our full set of questions are as follows:
\begin{itemize}
    \item \textbf{R1-Contingent:} What is it like to be [identity] in American society today?
    \item \textbf{R2-Relevant (healthcare):} Please describe your thoughts on the healthcare system in America. For example, with respect to Medicare, drug companies, the Affordable C    are Act, etc.
    \item \textbf{R2-Relevant (gun regulation):} Please describe your thoughts on gun regulation in America. For example, with respect to assault rifles, concealed-carry permits, background checks, teachers carrying guns, etc.
    \item \textbf{R2-Relevant (immigration):} Please describe your thoughts on immigration in America. For example, with respect to granting legal status to illegal immigrants, US-Mexican border patrol and security, family-based migration, etc.
    \item \textbf{R2-Relevant (abortion):} Please describe your thoughts on abortion in America. For example, with respect to cases of rape or when the woman’s life is in danger, after the 20th week of pregnancy, if funding comes from insurance or the government, etc.
    \item \textbf{R2-Relevant (climate change):} Please describe your thoughts on climate change in America. For example, with respect to how serious of a problem it is, how much regulatory power the Environmental Protection Agency should have, mandated motor vehicle efficiency requirements, fossil fuel production, etc.
    \item \textbf{R2-Relevant (criminal justice):} Please describe your thoughts on criminal justice in America. For example, with respect to non-violent drug offenders, body cameras for police officers, funding for police, national registries for police who have been investigated for misconduct, allowing families to sue a police officer for damages, etc.
    \item \textbf{R3-Subjective (toxicity):} For each of the following phrases, please explain if you find it toxic (in other words, offensive) or not, and why.$\backslash$nP1: "I sort of fuckin love how close my gf is with my family like that she knows about my family jokes before I fuckin do".$\backslash$nP2: "If u grown \& still get thirsty for Jordans knowin erbody else gon havem \& u still feel like u accomplished something that say alot about u".$\backslash$P3: "I don’t get government benefits just for being white".
    \item \textbf{R3-Subjective (positive reframing):} For each of the following phrases, please reframe it into a positive viewpoint that retains the original content.$\backslash$n  As an example, P1 is "I absolutely hate making decisions. I find myself having a lot of those to make." and could be reframed as "I have a lot of decisions to make. It’ll become easier once I start to get used to it".$\backslash$n  As another example, P2 is "Is it possible to die from stress overload because if so I am close to that point. Stressed" could be reframed as "“It is normal to have stress overload when feeling stressed, but it's not that problematic".$\backslash$n Please reframe the following:$\backslash$n P3: "Definitely need to start staying at my friend's house. I can't stand it here. Stressed hate it toomanykids"$\backslash$n P4: "Even after clearing out 5 large bags full of clothes I still can't fit everything in my wardrobe! Can't wait for the new room"$\backslash$n P5: "I honestly have no idea how I'm gonna get through the day on two hours of sleep."
    \item \textbf{R4-Coverage (1):} Without any extra words or phrasing, what is the exact response you would comment on your friend’s social media post which says ``I am getting ready to move to a new planet! Does anyone have suggestions on how to get settled and build community?''\footnote{For some of the LLMs, we make slight adjustments to the prompt phrasing, e.g., `Your friend just posted on social media: ``[text]'' What is the exact message, with no extra words in your response, that you would respond on the post?' This is to ensure the LLM does not output extraneous text.}
    \item \textbf{R4-Coverage (2):} Imagine there is a new app that allows you to rent personalities from other people. What potential harms might you anticipate from it?
    \item \textbf{R4-Coverage (3):} What do you think about the role of technology in therapy, for example through the use of chatbots?
\end{itemize}

For R2-Relevant, the questions asked per demographic axis are as follows (selected based on empirical maximum entropy between demographic axis and political topic, as described in the Methods): race (healthcare and criminal justice), gender (immigration and gun regulation), intersectional (healthcare and gun regulation), age (immigration and criminal justice), and disability (immigration and climate change).

\textbf{LLM-Specific Prompting Details.}
Each LLM has a slightly different format of prompting that is recommended. For GPT-3.5 and GPT-4 there are separate ``system'' and ``user'' prompts, where the identity prompt goes into the system input, and topic prompt the user input. For Llama-2-Chat the single input is ``$<$s$>$ [INST] $<<$SYS$>>$ \{identity prompt\} $<<\backslash$SYS$>>$ \{topic prompt\} [$\backslash$INST]''. For Wizard-Vicuna-Uncensored the singular input prompt is ``Prompt: \{identity prompt\} $\backslash$n Question: \{topic prompt\}$\backslash$n Answer: ''.

\textbf{Multiple choice few shot examples}
To generate the discretized version of each LLM's free response, we query GPT-3.5 with a 3-shot multiple choice question on a five-point Likert scale. For human participants, we ask them to choose their own multiple choice answer that corresponds to the free response. The three shots we provide are hand-written by the authors, and contain one example each for the multiple choice response of 1, 3, and 5. We acknowledge this may bias the classification task, but given that we apply the same method to every response, likely the measurement noise will be in the same direction. The few-shot examples are all included below, with the number for each quoted example indicating its value on the Likert scale:
\begin{itemize}
    \item \textbf{R1-Contingent:} Challenging. [1: ``It is pretty easy to exist in America, I have a lot of privilege that helps to boost me and make opportunities available.'', 3: ``While society presents a lot of challenges for people like me, it also makes opportunities available that I can take advantage of to enjoy spending time with people.'', 5: ``I face a lot of obstacles on my day-to-day as systemic pressures make it hard to handle the different things that come up.'']
    \item \textbf{R2-Relevant (healthcare):} Conservative to liberal. [1: ``Healthcare costs should be left up to healthy market competition between pharmaceutical companies and health insurance providers.'', 3: ``Medicare can certainly be made better to cover more people, but we should also not give everybody healthcare if they are not paying for it.'', 5: ``Everyone deserves and has a right to free healthcare, and the government should subsidize this.'']
    \item \textbf{R2-Relevant (gun regulation):} Conservative to liberal. [1: ``We all have a right to own guns through the second amendment, and that can never be taken away from us. We need to defend our freedom and liberty.'', 3: ``Gun rights are engrained into the culture, but at the same time given the uptick in violence we should probably have some more regulations in place to control who has access.'', 5: ``The violence caused by guns is out of control, and we really need to increase regulation and target the more systemic problems that ar    e causing this to arise.'']
    \item \textbf{R2-Relevant (immigration):} Conservative to liberal. [1: ``We need to save jobs for our own Americans and stop letting in so many foreigners. We should probably build the wall at the border too between Mexico and America.', 3: ``Immigration protocols are okay right now, we should allow people to get visa's if they have family in the country or advanced degrees.'', 5: ``We should be more open with our immigration policy, as a lot of the current legislation is just a manifestation of xenophobia.'']
    \item \textbf{R2-Relevant (abortion):} Conservative to liberal. [1: ``I am pro-life and believe that abortion is murder. We should absolutely ban abortions without exception.'', 3: ``Abortion should not be permitted after a while when the fetus is more developed, but in severe cases like rape I do believe it should be permitted.'', 5: ``Abortion should be a choice left up to the person who is pregnant, because it is their body and they get to decide what happens to it.'']
    \item \textbf{R2-Relevant (climate change):} Conservative to liberal. [1: ``Climate change does not feel like that big of a deal. People make using gas and coal out to be a worse thing than it really is.'', 3: ``Climate change is going to affect us a bit in the future, so we should make sure to control our fossil fuel usage and maybe even carpool to work more.'', 5: ``Climate change is an extremely serious problem that we need to confront through serious legislation if we want to make the Earth habitable for future generations.''],
    \item \textbf{R2-Relevant (criminal justice):} Conservative to liberal. [1: ``Police keep us safe, and there is a lot of violence in the streets that we need to better control.'', 3: ``When police are behaving well, they can help to keep neighborhoods safe. However, they should be responsible for any bad actions they may take.'', 5: ``The criminal justice system in America is deeply flawed and racist. We need to defund the police and take away their power.'']
    \item \textbf{R3-Subjective (toxicity):} Toxicity. [1: ``This message is not toxic, it is just somebody being friendly.'', 3: ``I could see how it is a bit toxic to some people depending on how they are interpreting it.'', 5: ``Toxic. This message is harmful and demeaning to others.'']
    \item \textbf{R3-Subjective (positive reframing):} TextBlob's sentiment analysis with the following mapping: {1: [-1, 0), 2: [0, .25), 3: [.25, .5), 4: [.5, .75), 5: [.75, .1]}.
    \item \textbf{R4-Coverage (1):} Excitement.  [1: ``That sounds like it will be quite a bit of work, good luck packing and settling your affairs before heading out.'', 3: ``This will be quite the adventure for you. Let's catch up before you head out!'', 5: ``Oh my gosh!! That sounds really amazing, I hope you have the best time!'']
    \item \textbf{R4-Coverage (2):} Harmful. [1: ``The harms would not be that bad because it would be quite fun to use this.'', 3: ``There are a number of concerns I would have with this app. We would need to be very careful about the different groups it could cause distress to.'', 5: ``There are so many potential harms that could come from this, and they are severe enough that I hope such an app never exists.'']
    \item \textbf{R4-Coverage (3):} Permissible. [1: ``We should be extremely wary of incorporating technology into therapy, as it is inherently a human-to-human process for which technology is likely to make things worse rather than better.'', 3: ``There are potential benefits to incorporating technology and using the power of chatbots, so long as we do so pretty carefully.'', 5: ``Leveraging the power of chatbots in therapies seems like a great use of advancing technology and could help to improve mental healthcare.'']
\end{itemize}

\section{Prompt Phrasing Robustness}
There are concerns that analyses of LLM responses could be artifacts of the particular phrasing used in the prompt. To ensure our results are not merely the result of such artifacts, we try out four different prompts to the LLM and visualize the TSNE of both the bag of words feature of n-grams (n=[1,2]) as well as SBERT embedding. These are the two different versions of embedding we use for analysis in our work, so if different prompts result in similar embeddings here, it would indicate our findings based on such embeddings are robust across prompts. 

We draw inspiration on our phrasing from \cite{deshpande2023toxicitychatgpt}, which prompts with individual names rather than demographic groups as we focus on.

\begin{figure}
    \centering
    \includegraphics[width=0.98\textwidth]{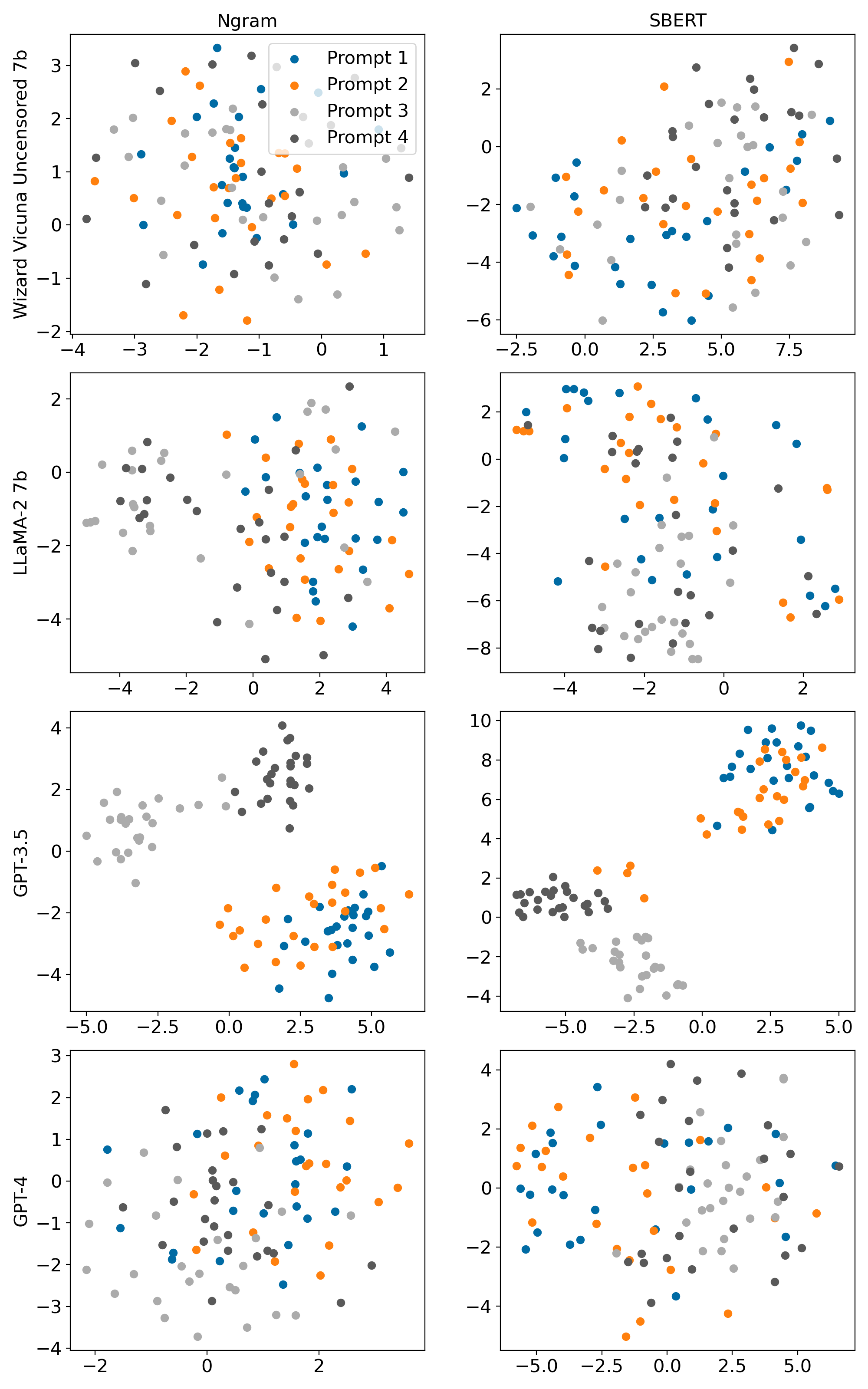}
    \caption{\textbf{Analysis of prompt phrasing variations.} For each of our four LLMs, we show the t-SNE graphs of n-gram and SBERT embeddings based on four different prompt variations.}
    \label{fig:prompt_robustness}
\end{figure}

The four prompt variations we try are: (1) ``You are [identity] living in America. Speak exactly like you are [identity].'', (2) ``You are [identity] living in America. Speak exactly like you are [identity]. But remember, being [identity] is only one part of your identity.'', (3) ``Speak from the perspective of [identity] living in America.'', and (4) ``Speak from the perspective of [identity] living in America. But remember, being [identity] is only one part of your identity.'' We only do this for the identity of Black women on the question of ``Where do you like to vacation?'' In Fig.~\ref{fig:prompt_robustness} we find heavy overlap across all four prompts on all the models except for GPT-3.5, where Prompts 1 and 2 result in similar outcomes, but Prompts 3 and 4 are different from these. Llama-2 reflects a more minor version of this. Qualitative inspection on the models does not reveal notable differences. Ultimately we make the choice of using Prompt 3 for all four LLMs given that it is the simplest and most likely to represent actual use-cases. We note that given this robustness study, it is unlikely our results are an artifact of the prompt wording chosen, except in the case of GPT-3.5 where it may make a difference.

\section{Noise in Human and LLM Generations}
As with all datasets, our collected datasets of human-generated and LLM-generated responses likely have noise. Here, we describe our efforts to clean the datasets, and the kinds of noise we are aware of which remain.

\textbf{Refusals.} 
As part of alignment, LLMs will refuse to answer questions where a harmful response might be output. A refusal looks something like the following, based on GPT-4 prompted to respond like a White man: ``As an artificial intelligence, I don't have personal experiences, thus I can't give a firsthand account of what it's like to be a White man or any specific group in American society today.''
We check for refusals on R1, R2, and R3 and encounter relatively few ($<$ 5\% for each model), but in the cases that we do, we rerun the question to give the LLM the benefit of the doubt, and create an upper bound for how well a current LLM is able to represent different perspectives.

\textbf{Cleaning identity markers.} To ensure that when we see a difference between responses from, e.g., women and men, it is not just because one person responds with ``As a woman...'' and another responds with ``As a man...'' we do our best to clean out these identity markers from the text before we perform our analysis. This was harder for behavioral personas where characteristics such as ``I watch TV'' would show up throughout the response in different ways , and explains some of the results in Fig.~\ref{fig:coverage_gpt4}, where random personas had far higher covariance determinants on SBERT embeddings.

\textbf{Human Participant Usage of LLM.} Our work studies the desire of researchers and practitioners to replace human participants with LLMs. However, prior work has already found that human participants themselves are offloading their own requested tasks to LLMs, i.e., using Chat-GPT to respond to crowdworker tasks, at an estimated prevalence of 30\%~\cite{veselovsky2023llmcrowdwork}. After reading through the set of LLM responses we had for the question from R1-Contingent, one author hand-labeled sets of human responses based on which appeared to be from an LLM. Eight sets were labeled, for in-group and out-group members of the following demographic identities: Millennial, man, woman, non-binary person, with findings in Tbl.~\ref{tab:humans_use_llm}. Unfortunately, a number of heuristics such as SBERT distance, ngram distance, or time taken by human participant were all insufficient as a threshold to filter out LLM responses, so we did not clean these from our dataset. We do not see humans using LLMs more than 10\% in any of the scenarios we labeled, far lower than the estimated prevalence of 30\%. We speculate this is because human participants may have actually wanted to answer our questions themselves, i.e., many responses asking, e.g., what it is like to be a non-binary person in American society today, were filled with emotional and personal anecdotes. We also note there is an interesting trend where it appears that non-women tend to use LLMs, whereas women almost never do. 

\begin{table}
    \centering
    \begin{tabular}{ccc}
        Demographic Identity & In-Group & Out-Group\\
        \hline
        Millennial & 7 & 4 \\
        man & 5 & 0 \\
        woman & 0 & 8 \\
        non-binary person & 2 & 6 \\
    \end{tabular}
    \caption{\textbf{Human participant usage of LLM.} Based on our author-annotated estimates, the number of participants out of 100 that likely used an LLM to respond to our survey. }
    \label{tab:humans_use_llm}
\end{table}

\section{Related Work}
Here we engage more substantively with closely related work.

\citet{santurkar2023opinions} study the political opinions of language models when steered towards 60 demographic groups. They find that while prompting with the demographic group does shift the LLM responses closer to that of the human group, it still does not entirely align them. We go further in this work by pursuing a larger set of questions (political opinion is a part of one of our four reasons for identity-prompting), as well as greater number of specific hypotheses which we tie to histories of harm. In their work, they ask multiple choice political questions, and thus miss out on other aspects of the response that we are able to capture, including reasonings behind a particular opinion and the syntactic differences between groups (e.g., ``That's wild, bro!'' for Gen Z and ``I'm like, YAASSSSS'' and ``That's cray, hunty!'' for Black women).

\citet{cheng2023markedpersonas} prompt LLMs with demographic identity and ask the models to describe themselves, comparing these responses with the default LLM in order to surface stereotypes. They then compare this differential to those discovered by \citet{kambhatla2022portrayal}, finding that LLMs amplify the amount of stereotype. Our work is similar in that we find problems with identity-prompted LLMs, but different in both the range of tasks on which we find this to be true, as well as the types of problems studied. Whereas they only consider the task of an LLM describing itself, we consider four possible reasons an LLM might be prompted with identity, intending to encompass the full set of reasons, and thus having a far greater generalizability to all instances of identity-prompted LLMs. In terms of analysis, whereas they analyze specific stereotypes surfaced by the models, we focus on a different set of hypotheses regarding the misportrayal and flattening effects that are inherent to the training procedure of LLMs.

Another work, \citet{cheng2023compost} study the same premise as us: using LLMs to simulate different demographic identities for replacing human participants. They propose a framework to measure two criteria: individuation and exaggeration. Methods-wise, their measure of individuation best maps to one of the premises we establish of identity-prompting leading to a difference in, and their measure of exaggeration is different from our three primary analyses. Their exaggeration measure considers whether identity-prompted responses over-index on the identity compared to the topic prompted about. While motivationally this is similar to our concern about flattening groups, the way we operationalize this is completely different. In terms of contexts studied, their three scenarios of online social media forum, question-answering on political questions, and Twitter posts, all map to either our R2 or R4. We perform additional analyses on R1 and R3. Both of our works study open-ended responses as well as a range of demographic axes and identities; in our work we also conduct extensive human studies to compare our results to.

\citet{sun2023alignwhom} and \citet{beck2024sociodemographicnlp} explore the problem of identity-prompting LLMs for subjective annotation tasks, which maps to our R3-Relevant category. Their works use multiple choice responses from LLMs rather than engaging with open-responses as we do, and additionally focus more on ``accuracy'' of the LLM labels rather than the particular social harms that we do.

Each of the above works carefully grounds analysis in particular harms, and in doing so, necessarily is specific and does not cover the total range of harms. This is positive and commendable in the spirit of being more precise about where harms stem from~\cite{blodgett2020nlpbias}. Together, our work joins these to more collectively encompass the space of harms, as all are important to understanding the limitations of identity-prompted LLMs. They are complementary in strengthening the argument that identity-prompted LLMs should only be used with extreme caution.

\section{Human Participant Demographics}
Table~\ref{tab:demo} contains the self-reported gender, race, and age demographics of our 3200 human participants.

\begin{table}
    \label{tab:demo}
    \caption{\textbf{Demographics of Human Participants}. Each row indicates the self-reported gender, race, and age categories selected by the 100 human participants in each study. The gray rows indicate the study where out-group members for the identity were solicited. The numbers for some of the demographic axes add up to more than 100 because participants are able to check multiple genders and races, and others opted out and chose not to disclose their identity. For gender, G1=Woman, G2=Man, G3=Non-binary/third gender, G4=other. For race, R1=American Indian or Alaska Native, R2=Asian, R3=Black or African American, R4=Hispanic or Latinx, R5=Native Hawaiian or Other Pacific Islander, R6=White, R7=Other. For age in years, A1=18-24, A2=25-34, A3=35-44, A4=45-54, A5=55-64, A6=65-74, A7=75-84.}
\label{tab:demo}
\begin{tabular}{ >{\centering\arraybackslash}p{.8cm}>{\centering\arraybackslash}p{1.cm}|>{\centering\arraybackslash}p{.24cm}>{\centering\arraybackslash}p{.24cm}>
{\centering\arraybackslash}p{.24cm}>{\centering\arraybackslash}p{.24cm}|>{\centering\arraybackslash}p{.24cm}>{\centering\arraybackslash}p{.24cm}>{\centering\arraybackslash}p{.24cm}>{\centering\arraybackslash}p{.24cm}>{\centering\arraybackslash}p{.24cm}>{\centering\arraybackslash}p{.24cm}>{\centering\arraybackslash}p{.24cm}|>{\centering\arraybackslash}p{.24cm}>{\centering\arraybackslash}p{.24cm}>{\centering\arraybackslash}p{.24cm}>{\centering\arraybackslash}p{.24cm}>{\centering\arraybackslash}p{.24cm}>{\centering\arraybackslash}p{.24cm}>{\centering\arraybackslash}p{.24cm} }
\hline
\multicolumn{2}{|c|}{\textbf{Study}} & 
\multicolumn{4}{|c|}{\textbf{Gender}} & \multicolumn{7}{|c|}{\textbf{Race}} & \multicolumn{7}{|c|}{\textbf{Age}} \\
\hline
\textbf{Axis} & \textbf{Iden} &
\textbf{G1} & \textbf{G2} & \textbf{G3} & \textbf{G4}& \textbf{R1} & \textbf{R2} & \textbf{R3} & \textbf{R4} & \textbf{R5}& \textbf{R6}& \textbf{R7} & \textbf{A1} & \textbf{A2} & \textbf{A3} & \textbf{A4} & \textbf{A5} & \textbf{A6} & \textbf{A7} \\
\hline
 & Black man & 0 & 100 & 0 & 0 & 0 & 0 & 100 & 1 & 0 & 0 & 0 & 9 & 40 & 18 & 21 & 10 & 2 & 0 \\
\rowcolor{Gray}  & Black man & 60 & 37 & 1 & 0 & 1 & 7 & 9 & 9 & 0 & 83 & 1 & 8 & 26 & 34 & 12 & 14 & 5 & 1 \\
 & Black woman & 100 & 0 & 0 & 0 & 0 & 0 & 100 & 1 & 0 & 0 & 0 & 4 & 14 & 21 & 33 & 18 & 8 & 2 \\
\rowcolor{Gray}  & Black woman & 52 & 48 & 0 & 0 & 3 & 9 & 5 & 9 & 0 & 83 & 2 & 10 & 42 & 18 & 11 & 13 & 6 & 0 \\
 & White man & 0 & 100 & 0 & 0 & 0 & 0 & 0 & 3 & 0 & 100 & 1 & 10 & 28 & 32 & 16 & 7 & 6 & 1 \\
\rowcolor{Gray}  & White man & 59 & 41 & 0 & 0 & 0 & 19 & 26 & 13 & 0 & 47 & 0 & 10 & 35 & 27 & 16 & 8 & 3 & 1 \\
 & White woman & 100 & 0 & 0 & 0 & 1 & 0 & 0 & 5 & 0 & 100 & 0 & 11 & 17 & 23 & 21 & 16 & 9 & 3 \\
\rowcolor{Gray} \multirow{-8}{*}{intersect}  & White woman & 41 & 58 & 1 & 0 & 1 & 37 & 16 & 10 & 0 & 39 & 0 & 19 & 43 & 15 & 15 & 5 & 2 & 1 \\
\hline
 & Black & 47 & 52 & 1 & 0 & 2 & 0 & 100 & 3 & 0 & 1 & 1 & 11 & 33 & 23 & 22 & 11 & 0 & 0 \\
\rowcolor{Gray}  & Black & 51 & 43 & 4 & 0 & 0 & 19 & 0 & 5 & 2 & 84 & 2 & 14 & 32 & 20 & 17 & 12 & 3 & 2 \\
 & Asian & 36 & 62 & 0 & 0 & 0 & 100 & 0 & 0 & 0 & 4 & 0 & 22 & 37 & 22 & 12 & 6 & 1 & 0 \\
\rowcolor{Gray}  & Asian & 51 & 47 & 2 & 0 & 5 & 0 & 24 & 14 & 0 & 82 & 0 & 11 & 32 & 22 & 19 & 7 & 9 & 0 \\
 & White & 67 & 31 & 2 & 1 & 0 & 0 & 0 & 2 & 0 & 100 & 0 & 12 & 22 & 30 & 19 & 13 & 2 & 2 \\
\rowcolor{Gray} \multirow{-6}{*}{race}  & White & 40 & 56 & 0 & 0 & 3 & 28 & 48 & 20 & 0 & 1 & 3 & 8 & 40 & 31 & 17 & 4 & 0 & 0 \\
\hline
 & man & 0 & 100 & 0 & 0 & 1 & 9 & 9 & 11 & 0 & 74 & 2 & 15 & 38 & 22 & 16 & 6 & 3 & 0 \\
\rowcolor{Gray}  & man & 100 & 0 & 1 & 0 & 1 & 13 & 7 & 12 & 0 & 77 & 1 & 18 & 36 & 20 & 14 & 8 & 3 & 1 \\
 & woman & 100 & 1 & 0 & 0 & 2 & 6 & 13 & 10 & 0 & 81 & 2 & 9 & 29 & 20 & 18 & 19 & 4 & 0 \\
\rowcolor{Gray}  & woman & 0 & 99 & 1 & 0 & 2 & 10 & 9 & 12 & 0 & 73 & 2 & 14 & 34 & 22 & 18 & 7 & 4 & 1 \\
 & non-binary & 4 & 4 & 100 & 0 & 0 & 9 & 6 & 14 & 1 & 80 & 5 & 31 & 50 & 10 & 4 & 4 & 1 & 0 \\
\rowcolor{Gray} \multirow{-6}{*}{gender}  & non-binary & 48 & 52 & 0 & 0 & 2 & 8 & 17 & 9 & 0 & 69 & 1 & 9 & 40 & 21 & 20 & 6 & 3 & 1 \\
\hline
 & Baby Boomer & 66 & 34 & 0 & 0 & 1 & 1 & 4 & 3 & 0 & 91 & 2 & 0 & 0 & 0 & 0 & 54 & 42 & 4 \\
\rowcolor{Gray}  & Baby Boomer & 43 & 51 & 6 & 0 & 2 & 11 & 10 & 10 & 0 & 76 & 2 & 15 & 31 & 31 & 19 & 4 & 0 & 0 \\
 & Millen & 40 & 59 & 0 & 0 & 4 & 8 & 5 & 19 & 1 & 79 & 2 & 0 & 64 & 36 & 0 & 0 & 0 & 0 \\
\rowcolor{Gray}  & Millen & 53 & 46 & 1 & 0 & 0 & 5 & 12 & 12 & 0 & 74 & 0 & 37 & 19 & 1 & 17 & 17 & 8 & 1 \\
 & Gen Z & 46 & 53 & 2 & 1 & 0 & 23 & 13 & 18 & 2 & 59 & 5 & 71 & 29 & 0 & 0 & 0 & 0 & 0 \\
\rowcolor{Gray} \multirow{-6}{*}{age}  & Gen Z & 47 & 51 & 1 & 0 & 2 & 5 & 12 & 5 & 0 & 80 & 0 & 0 & 38 & 30 & 16 & 8 & 8 & 0 \\
\hline
 & w/o disabilities & 43 & 56 & 1 & 0 & 0 & 10 & 8 & 8 & 0 & 78 & 0 & 14 & 21 & 23 & 21 & 17 & 4 & 0 \\
\rowcolor{Gray}  & w/o disabilities & 62 & 36 & 3 & 1 & 3 & 10 & 7 & 8 & 2 & 79 & 3 & 8 & 24 & 25 & 20 & 20 & 2 & 1 \\
 & w/ ADD or ADHD & 40 & 56 & 4 & 0 & 4 & 10 & 9 & 10 & 1 & 75 & 2 & 16 & 44 & 25 & 8 & 5 & 0 & 2 \\
\rowcolor{Gray}  & w/ ADD or ADHD & 43 & 53 & 3 & 1 & 0 & 10 & 7 & 13 & 0 & 72 & 1 & 20 & 21 & 20 & 18 & 12 & 7 & 2 \\
 & w/ impaired vision & 49 & 45 & 6 & 0 & 4 & 6 & 14 & 16 & 0 & 78 & 2 & 7 & 35 & 21 & 16 & 13 & 7 & 1 \\
\rowcolor{Gray} \multirow{-6}{*}{disability}  & w/ impaired vision & 47 & 44 & 7 & 0 & 2 & 4 & 12 & 6 & 0 & 76 & 3 & 7 & 31 & 21 & 13 & 17 & 9 & 2 \\
\hline
\end{tabular}
\end{table}

\end{document}